\documentclass[reprint, superscriptaddress, amsmath, amssymb, aps, prl]{revtex4-2}

\usepackage{graphicx}
\usepackage{dcolumn}
\usepackage{bm}

\usepackage[utf8]{inputenc}
\usepackage{amsmath}
\usepackage{amsfonts}
\usepackage{amssymb}
\usepackage[export]{adjustbox}
\usepackage[svgnames,dvipsnames]{xcolor}
\usepackage{geometry}

\usepackage{caption}
\usepackage{subcaption}
\usepackage{tikz}
\usepackage{gensymb}
\usetikzlibrary{patterns, arrows.meta, calc, decorations.pathreplacing}
\definecolor{waterblue}{RGB}{180, 220, 255} 

\usepackage{natbib} 
\usepackage{amssymb}
\usepackage{amsthm}
\usepackage{epstopdf, epsfig}

\usepackage{placeins}
\usepackage{pgfplots}
\usepgfplotslibrary{fillbetween}
\pgfplotsset{compat=1.15}
\newlength\figH
\newlength\figW
\usepackage{overpic} 
\usepackage{tabularx}  

\usepackage{array}     
\usepackage{mathtools}  
\usepackage{adjustbox} 
\pgfplotsset{compat=newest}

\definecolor{cadmiumorange}{rgb}{0.93, 0.53, 0.18}

\usepackage{hyphenat}

\newcommand{\We}{\mathrm{We}}
\renewcommand{\Re}{\mathrm{Re}} 
\newcommand{\Oh}{\mathrm{Oh}}

\begin{document}

\preprint{APS/123-QED}

\title{Kinematic Closure of Drop Impact}

\author{Mete Abbot}
 \email{m.abbot@uva.nl}
 \affiliation{Van der Waals-Zeeman Institute, Institute of Physics, University of Amsterdam, Science Park 904, 1098 XH Amsterdam, Netherlands}
\affiliation{Department of Aerospace Engineering, Faculty of Engineering, Dokuz Eyl\"ul University, 35390 Izmir, Turkey}

\author{Daniel Bonn}
 \email{d.bonn@uva.nl}
 \affiliation{Van der Waals-Zeeman Institute, Institute of Physics, University of Amsterdam, Science Park 904, 1098 XH Amsterdam, Netherlands}

\date{\today}

\begin{abstract}
Predicting how far a liquid droplet spreads upon impacting a solid surface 
is governed by a complex competition between inertia, surface tension, and 
viscous dissipation. Existing models typically estimate this maximum spreading diameter by prescribing the spreading time or velocity from a decoupled asymptotic limit, restricting predictive accuracy across regimes. Here, we formulate a reduced kinematic description by resolving both the characteristic spreading time and spreading velocity from an energy balance coupling capillary storage and viscous dissipation. 
Unifying these two asymptotically constrained closures via the kinematic 
relation $D_{\max} = V_s t_s$ yields a compact, parameter-free approximation for the maximum spreading ratio that bridges the 
inertio-capillary and inertio-viscous regimes. We quantify the accuracy of 
this reduced closure against the exact implicit energy relation and demonstrate that it quantitatively collapses our experimental measurements alongside extensive literature data across wide ranges of Weber and Ohnesorge numbers, droplet sizes, and surface wettabilities without adjustable prefactors.
\end{abstract}

\maketitle


The impact of drops on solid surfaces is a key step in processes across vastly different scales: from the environmental transport of rainfall and the deposition of respiratory aerosols, to the precision of ink-jet printing and pesticide distribution \cite{josserand2016drop, yarin2006drop,cheng2022drop}. Despite over a century of study \cite{worthington1895splash}, establishing a universal prediction for the maximum spreading diameter $D_\mathrm{max}$ of a droplet of initial diameter $D_0$ impacting on a solid surface remains an open problem due to the vast possible parameter space (shown in Fig.~\ref{fig:parameter_space}; Vertical axis is the Ohnesorge number $\Oh = \eta/\sqrt{\rho \sigma D_0}$; horizontal axis is the Weber number $\We = \rho D_0 {V_0}^2/\sigma$; Background coloring is based on the Reynolds number $\Re = \rho D_0 V_0/\eta$, where $\eta$ is the dynamic viscosity, $\rho$ density, $\sigma$ surface tension, $D_0$ initial droplet diameter, and $V_0$ fluid velocity before impact). 

Energy conservation provides descriptions of the asymptotic inertio–capillary and inertio–viscous limits, leading to the well-known scalings $\beta_\mathrm{max} \equiv D_\mathrm{max}/D_0 \sim \We^{1/2}$ and $\beta_\mathrm{max} \sim \Re^{1/5}$ in their respective regimes, where $\We$ is the Weber number and $\Re$ the Reynolds number \cite{clanet2004maximal, eggers2010drop, roisman2009inertia, gordillo2019theory}. Building on these limits, several authors have proposed “bridging” frameworks that allow collapsing data, most prominently the Padé-type interpolation in terms of the impact parameter $P = \We\,\Re^{-2/5}$ introduced by Laan \textit{et al.} \cite{laan2014maximum} and subsequently refined or reinterpreted in various contexts \cite{lee2016universal,de2019predicting}.

\begin{figure}[b]
    \centering
    \input{Figures/parameter_space_images}
    \caption{\textbf{Droplet morphology at maximum spreading.} Insets correspond to scaling limits: in regime I (low $\We$, low $\Oh$), $D_\mathrm{max}$ scales as $\We^{1/2}$ \cite{de2019predicting,lee2016universal,laan2014maximum}; in regimes II, III, and IV (high $\We$), as $\Re^{1/5}$ \cite{abbot2024spreading,jorgensen2024deformation,eggers2010drop,roisman2009inertia}; and in regime V (low $\We$, high $\Oh$), as $\Re^{1/3}$ \cite{jorgensen2024deformation}.}.
    \label{fig:parameter_space}
\end{figure}

\begin{figure}[t!]
    \centering
        \begin{subfigure}[b]{0.23\textwidth}
        \centering
        \input{Figures/Laanrepresentation} 
        \caption{}
        \label{fig:Laan}
    \end{subfigure}
    \hfill
    \begin{subfigure}[b]{0.23\textwidth}
        \centering
        \pgfplotsset{compat=1.17} 

\definecolor{mycolor0}{RGB}{50,155,105}    
\definecolor{mycolor5}{RGB}{213,81,156}    
\definecolor{mycolor10}{RGB}{54,80,163}    
\definecolor{mycolor125}{RGB}{244,196,48}  
\definecolor{mycolor15}{RGB}{245,127,34}   
\definecolor{mycolorSplash}{RGB}{204,204,255} 
\definecolor{mycolorNo}{RGB}{224,224,224}     

\begin{tikzpicture}
\begin{axis}[
width=0.8\figW,
height=0.8\figH,
at={(0\figW,0\figH)},
    xlabel={\footnotesize $t\cdot V_0/D_{\mathrm{max}}$},
    ylabel={\footnotesize $D_{(t)}/D_{\mathrm{max}}$},
    xmin=0, xmax=0.7,
    ymin=0, ymax=1.2,
    ytick={0, 1},
    legend style={at={(0.15,0.99)}, anchor=north west, draw=none, font=\scriptsize},
    ticklabel style = {font=\scriptsize}, legend cell align=left
    ]

    \addplot graphics[
        xmin=0.025, xmax=0.7,  
        ymin=0.03, ymax=1.05   
    ] {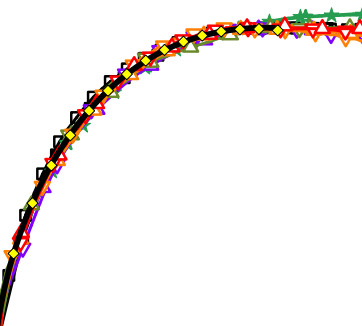};

\addplot[dashed]
table[row sep=crcr]{
0.55 0 \\
0.55 1 \\
};

\addplot[dashed]
table[row sep=crcr]{
0 1 \\
0.5 1 \\
};

\addplot[color=black, line width=0.5pt]
table[row sep=crcr]{
0.7 0 \\
0.7 2 \\
};

\node[anchor=south east] at (rel axis cs: 0.76, 0.15) {\scriptsize Exp. \cite{gorin2022universal}};
\node[anchor=south east] at (rel axis cs: 0.76, 0.07) {\scriptsize \textbf{--} Theory \cite{gordillo2019theory}};

\end{axis}
\end{tikzpicture} 
        \caption{}
        \label{fig:Gorin}
    \end{subfigure}
    
    
    \begin{subfigure}[b]{0.23\textwidth}
        \centering
        \pgfplotsset{compat=1.17} 

\definecolor{mycolor0}{RGB}{50,155,105}    
\definecolor{mycolor5}{RGB}{213,81,156}    
\definecolor{mycolor10}{RGB}{54,80,163}    
\definecolor{mycolor125}{RGB}{244,196,48}  
\definecolor{mycolor15}{RGB}{245,127,34}   
\definecolor{mycolorSplash}{RGB}{204,204,255} 
\definecolor{mycolorNo}{RGB}{224,224,224}     

\begin{tikzpicture}
\begin{axis}[
width=0.8\figW,
height=0.8\figH,
at={(0\figW,0\figH)},
    xlabel={\footnotesize Re},
    ylabel={\footnotesize $t_s/(D_0/V_0)$},
    xmin=0.1, xmax=100000,
    ymin=0, ymax=6,
    xmode=log,
    minor x tick num=1,  
    minor y tick num=1,  
    legend style={at={(0.01,0.99)}, anchor=north west, draw=none, font=\scriptsize},
    ticklabel style = {font=\scriptsize}, legend cell align=left
    ]

\addplot[only marks, color=blue, mark=square, mark size=2.0pt, line width=0.5pt]
table[row sep=crcr]{
1036.377696	0.794154972	\\
1963.220851	1.636008761	\\
2399.097238	1.14898688	\\
3687.178199	1.553974967	\\
5071.061495	2.42865651	\\
7072.570968	2.845272189	\\
9816.750475	4.607454411	\\
11913.09295	4.450269911	\\
13476.30271	4.776059144	\\
};
\addlegendentry{Water}

\addplot[only marks, color=violet, mark=triangle, mark size=2.0pt, line width=0.5pt]
table[row sep=crcr]{
93.95849031	0.723801087	\\
178.9295123	0.931854416	\\
283.6544019	1.446479348	\\
356.9927625	1.588064437	\\
492.5627673	1.817043605	\\
683.0940029	2.223445964	\\
946.4943739	1.437708778	\\
1142.207713	2.726419135	\\
1314.635104	2.139544559	\\
};
\addlegendentry{Glycerin 60\%}

\addplot[only marks, color=red, mark=o, mark size=2.0pt, line width=0.5pt]
table[row sep=crcr]{
0.555655999	0.21048011	\\
1.46640418	0.40223517	\\
2.380113705	0.590688291	\\
3.017745117	0.630680778	\\
4.261644606	0.779313718	\\
5.963251365	0.701023669	\\
8.261075576	0.971149656	\\
9.937555772	1.168232125	\\
11.44550852	1.196002634	\\
12.53815507	1.146406746	\\
};
\addlegendentry{Glycerin}

\addplot[dashed, color=black, line width=2.0pt]
table[row sep=crcr]{
0.1	1	\\
100000	1	\\
};

\end{axis}
\end{tikzpicture} 
        \caption{}
        \label{fig:timefail}
    \end{subfigure}
    \hfill
    \begin{subfigure}[b]{0.23\textwidth}
        \centering
        \begin{tikzpicture}
\begin{axis}[
width=0.8\figW,
height=0.8\figH,
at={(0\figW,0\figH)},
    xlabel={\footnotesize Re},
    ylabel={\footnotesize $V_s/V_0$},
    xmin=0.05, xmax=50000,
    ymin=0.7, ymax=4,
    xmode=log,
    minor x tick num=1,  
    minor y tick num=1,  
    legend style={at={(0.01,0.99)}, anchor=north west, draw=none, font=\scriptsize},
    ticklabel style = {font=\scriptsize}, legend cell align=left
    ]

    \addplot[
        scatter,
        only marks,
        mark=*,
        mark size=2pt,
        mark options={fill opacity=0.8},
        colormap={blue-red}{rgb255(0cm)=(0,0,255); rgb255(1cm)=(255,0,0)},
        scatter src=explicit,
        forget plot   
    ] table [x=Re, y=Ratio, meta=LogOh, col sep=comma] {Figures/Data/velocity_data_for_latex_Re.csv};

\node[anchor=south, font=\scriptsize, color=red] at (axis cs: 1, 3.0) {Oh$\sim 10^2$};

\node[anchor=south, font=\scriptsize, color=blue] at (axis cs: 1000, 2.7) {Oh$\sim 10^{-3}$};

    \end{axis}
\end{tikzpicture}
        \caption{}
        \label{fig:ucharvsRe}
    \end{subfigure}
    \caption{\textbf{Predictive conundrum.} (a) Maximum spreading ratio $\beta_{\mathrm{max}}$ based on the equatorial width, multiplied by $\Re^{-1/5}$ for better visibility, versus impact parameter $P=\We\,\Re^{-2/5}$ from the experiments of \cite{liu2025maximum,jorgensen2024deformation,mclauchlan2025bouncing}. The solid black line represents the empirical fit proposed in \cite{laan2014maximum}. (b) Temporal evolution data collapse suggests that a unified physical scaling exists \cite{gorin2022universal}.  (c) The experimental spreading time $t_s$ from \cite{liu2025maximum,aksoy2022spreading} does not correlate with the inertial timescale.  (d) The characteristic velocity $V_s$ deviates from $V_0$ (combined data from this work and \cite{aksoy2022spreading,liu2025maximum}).}
\end{figure}

However, $P$–bridging is not universally valid across the full $(\We,\Oh)$ space, as shown in Fig.~\ref{fig:Laan}. For large Ohnesorge numbers, experiments and numerical simulations show that viscous dissipation rapidly fills the drop volume, leading to a qualitatively new inertio–viscous regime \cite{liu2025maximum,sanjay2025unifying,jorgensen2024deformation} that is not captured by classical contact line dissipation theories \cite{gordillo2019theory, laan2014maximum}. At the other extreme of low impact energy (small $P$), even for low–viscosity liquids, experiments reveal that the “capillary” branch of the Padé fit does not represent a purely inertial–capillary balance: a fraction of the kinetic energy is dissipated during the short impact phase, modifying both the effective prefactor and the apparent scaling exponent relative to the ideal $\We^{1/2}$ law \cite{laan2014maximum,lee2016universal,abbot2024spreading}. While accounting for the finite zero-velocity spreading ratio $\beta_0$ removes some low-$P$ scatter \cite{lee2016universal, de2019predicting}, these discrepancies are clearly visible when $\beta_{\max} \Re^{-1/5}$ is plotted against $P$ for comprehensive datasets \cite{laan2014maximum,abbot2024spreading,aksoy2022spreading,jorgensen2024deformation,liu2025maximum}, as shown in Fig.~\ref{fig:Laan}.

In parallel with these developments, recent work has revealed a simple kinematic universality for droplet spreading. Gorin \textit{et al.} \cite{gorin2022universal} showed that the temporal evolution of the spreading diameter collapses onto a single curve when rescaled \emph{post-facto} by $D_{\mathrm{max}}$ and the impact velocity $V_0$ (Fig.~\ref{fig:Gorin}). Motivated by this observation, we posit that the maximum spreading ratio $\beta_{\mathrm{max}}$ is governed by the kinematic relation
\begin{equation}
    \beta_{\mathrm{max}} \equiv \frac{D_{\mathrm{max}}}{D_0}
    = \frac{V_s\, t_s}{D_0}.
    \label{eq:gorin_relation}
\end{equation}
We explicitly distinguish $V_s$ from the impact velocity $V_0$ to account for systematic deviations from inertial scaling ($V_s = V_0$).

Eq.~\eqref{eq:gorin_relation} is purely kinematic and valid regardless of the experimental conditions or liquid properties. The predictive difficulty arises because $V_s$ and $t_s$ are \emph{a priori} unknown dynamical outputs. Existing models predict spreading times representing asymptotic values of $t_s$ for specific regimes: inertial ($\tau_{i} \sim D_0/V_0$) \cite{roisman2009inertia,clanet2004maximal,gorin2022universal,abbot2024spreading,lagubeau2012spreading}, inertio–capillary ($\tau_{\mathrm{ic}} \sim \sqrt{\rho D_0^3/\sigma}$) \cite{liu2025maximum,sanjay2025unifying}, or viscous ($\tau_{v} \sim \rho D_0^2/\eta$) \cite{jorgensen2024deformation}. However, as shown in Fig.~\ref{fig:timefail}, detailed in \cite{aksoy2022spreading} and in the Supplementary Material S6, and discussed in detail below, measured spreading times do not follow any single asymptotic timescale across the parameter space; instead, they continuously interpolate between these limits.  Likewise, Fig.~\ref{fig:ucharvsRe} demonstrates that the characteristic spreading velocity $V_s$ systematically departs from $V_0$. Taking $t_{s}$ and $V_{s}$ from asymptotic limits inherently constrains models to discrete regimes, complicating the description of the continuous physics across the parameter space.

In this Letter, we resolve these constraints by treating the spreading time $t_s$ and the characteristic velocity $V_s$ not as predefined inputs, but as emergent dynamical quantities determined self-consistently from the energy balance. We formulate a closure from the energy balance for $t_s$ and $V_s$, explicitly retaining both capillary storage and viscous dissipation. This yields a unified spreading timescale that continuously bridges the inertial, viscous, and capillary regimes, and an associated velocity scale that captures the observed deviations from $V_s = V_0$. Substituting these self-consistent kinematic scales into Eq.~\eqref{eq:gorin_relation} leads to a unified scaling law for $\beta_{\mathrm{max}}$ that remains valid across a wide physical landscape, from inviscid to highly viscous liquids and from low to high impact energies.

We validate our predictions against $\sim$1,000 experiments and $\sim$1,400 simulations, combining the measurements in this work with a comprehensive dataset of the literature, expanding the data beyond the optical or time resolution limits of the literature. For the present measurements, a Phantom v7510 high-speed camera (temporal resolution $\sim$ 10 $\mu$s, spatial resolution $\sim 10~\mu\text{m/px}$) is used. Drops of water (1 mPa$\cdot$s, 72.3 mN/m), glycerol (1350 mPa$\cdot$s, 65 mN/m), and silicone oil (10 Pa$\cdot$s, 20 mN/m) are generated using a syringe pump (Harvard Apparatus) equipped with G21 needle stainless steel needle tip onto smooth borosilicate glass ($\theta=18^\circ \pm 5^\circ$, $<5$~nm roughness measured using a profilometer), with impact velocities of 0.1--5 m/s adjusted by changing the height. Our dataset combined with the literature dataset spans five orders of magnitude in viscosity ($10^{-3}$--$45$~Pa$\cdot$s), droplet sizes from $30~\mu\text{m}$ to $15~\text{mm}$, and contact angles from $0^\circ$ to $180^\circ$ \cite{jorgensen2024deformation,liu2025maximum,lee2016universal,laan2014maximum,du2021analytical,liu2023ionic,roisman2009inertia,cheng1977dynamic,pasandideh1996capillary,mclauchlan2025bouncing,marmanis1996scaling,naoe2014experimental,iqbal2025droplet,abbot2025nanoparticles}.

To determine the characteristic spreading time $t_s$, we formulate a closure from the energy balance ($E_k + E_{s,0} \approx E_{s,f} + W_{\nu}$). We explicitly retain the surface energy contribution ($E_{s,f}\sim \sigma D_{\mathrm{max}}^2$, assuming planar interfacial storage) to ensure a smooth transition to the thermodynamic wetting limit (see Supplementary Material S2). 

To evaluate the viscous work $W_{\nu}$, we model the dissipation through the lens of volume-limited lubrication flow. While early-time spreading involves a growing boundary layer \cite{lagubeau2012spreading,cheng2022drop,abbot2024spreading}, for highly viscous liquids ($\Oh \gtrsim 1$) and high deformations, this boundary layer rapidly fills the thickness of the drop. Crucially, both boundary-layer growth and lubrication-dominated dissipation yield an identical $\eta^{-1/5}$ scaling dependence on the spreading timescale. Therefore, representing the total effective dissipation via the volume-limited lubrication bound provides a physically bounded asymptotic closure. (See Supplementary Material S3 and S4 for the detailed derivation).

Under the assumptions of planar capillary interfacial storage ($E_{s,f} \sim \sigma D_{\max}^2$) 
and volume-limited lubrication shear ($h \sim D_0^3/D^2$), the central dimensionless energy relation becomes:
\begin{equation} \label{eq:master}
    \mathcal{E} \approx V^2 T^2 + \Lambda V^6 T^5,
\end{equation}
which represents a composite relation constrained by the inertio-capillary ($V^2 T^2$) and inertio-viscous ($\Lambda V^6 T^5$) asymptotic limits. $\mathcal{E} = 1 + 12/\We$ is the total dimensionless energy density, $T=t_s/\tau_{\mathrm{ic}} = t_s / \sqrt{\rho D_0^3 /\sigma}$ is the dimensionless time ratio to the inertio-capillary time, $V=V_s/V_0$ is the normalized velocity, and the unified damping parameter $\Lambda = \We^{2} \Oh$ emerges naturally from nondimensionalization.

Although Eq.~\eqref{eq:master} implicitly defines the spreading time, its physical behavior is best understood through its asymptotic limits. In the inviscid limit ($\Lambda \to 0$), the viscous term vanishes, yielding $T_{\mathrm{in}} \approx \sqrt{\mathcal{E}}/V$. In contrast, in the highly viscous limit ($\Lambda \gg 1$), dissipation dominates, resulting in a scaling of $T_{\mathrm{visc}} \approx (\Lambda V^6 \mathcal{E}^{-1})^{-1/5}$. 

Since the balance is a quintic equation and has no exact analytical solution, we capture the continuous transition by constructing a matched asymptotic interpolation ansatz using the harmonic mean ($T^{-1} \approx T_{\mathrm{in}}^{-1} + T_{\mathrm{visc}}^{-1}$) that recovers the correct limits while maintaining monotonicity:
\begin{equation}
        T \approx \frac{\sqrt{\mathcal{E}}}{V} \left[ 1 + (\Lambda V \mathcal{E}^{3/2})^{1/5} \right]^{-1}.
    \label{eq:time_analytical}
\end{equation}
While this is not the only possible interpolation that recovers correct asymptotes, comparison with the exact positive root reveals that this harmonic interpolation matches both asymptotic limits, with a maximum relative deviation bounded by $\approx 30\%$ in the crossover regime (near $\Lambda \sim 0.1$, see Supplementary Material Section S3). 

To formulate a predictive timescale decoupled from the spreading velocity, we evaluate the terms in Eq.~\eqref{eq:time_analytical} sequentially. First, the leading prefactor $\sqrt{\mathcal{E}}/V$ corresponds to the inertial limit. In this limit, the quadratic capillary energy $E_s \sim \sigma D^2$ yields a linear restoring force $F_c \sim \partial E_s/\partial D \sim \sigma D$ \cite{bartolo2005retraction}. Balancing $F_c \sim \sigma V_s t_s$ against the droplet's inertial deceleration $F_i \sim m(V_s/t_s)$ causes $V_s$ to cancel out identically, isolating the timescale from the velocity and yielding the constant geometric coefficient $T_0 = \sqrt{\pi/6}$.

Substituting $T_0$ in the viscous damping term yields $(\Lambda \mathcal{E}^2 / T_0)^{1/5} \approx (\Lambda \mathcal{E}^{2})^{1/5}$, since $T_0^{-1/5} \approx 1$. Further justification for this is provided in Supplementary Material S5. 

\begin{figure}[t!]
    \centering
    \begin{subfigure}[b]{\linewidth}
        \centering
        {\input{Figures/damping_time}}
        \caption{}
        \label{fig:time_closure}
    \end{subfigure}
    \\
    \begin{subfigure}[b]{\linewidth}
        \centering
        {\begin{tikzpicture}
    \begin{axis}[
        width=1.0\figW,
        height=0.9\figH, 
        xmode=log,     
        ymode=log, 
        xlabel={$\Lambda \mathcal{E}^2 / T$},
        ylabel={$V/V_{in}=(V_s/V_0)/(\sqrt{\mathcal{E}}/T)$},
        colorbar,
        colorbar style={
            ylabel={$\log_{10}(Oh)$},
            yticklabel style={/pgf/number format/fixed}
        },
        colormap={blue-red}{color(0cm)=(blue); color(1cm)=(red)},
        scatter/use mapped color={
            draw=black!50,      
            fill=mapped color   
        }, 
        ymin=0.006, ymax=3,     
        xmin=0.01, xmax=10^11,
        legend style={at={(0.02,0.25)}, anchor=north west, draw=none, fill=none, font=\footnotesize},
    ticklabel style = {font=\footnotesize}, legend cell align=left
    ]

    \addplot[
        scatter,
        only marks,
        mark=*,
        mark size=2pt,
        mark options={fill opacity=1.0}, 
        scatter src=explicit, 
        forget plot 
    ] table [x=y, y=x, meta=LogOh, col sep=comma] {Figures/Data/velocity_data_for_latex_Re_1.csv};

    \addplot[line width=2.0pt , domain=0.01:10^12, samples=200] 
    { ( (1 + x )^(-0.1667) };
    \addlegendentry{Eq.~\ref{eq:velocity_closure}}

    \coordinate (A) at (axis cs: 5*1e5, 0.05); 
    
    \coordinate (B) at (axis cs: 5*1e5, {0.05 * 10^(-1/6)}); 
    
    \coordinate (C) at (axis cs: 5*1e6, {0.05 * 10^(-1/6)}); 

    \draw[black, thick] (A) -- (B) node[midway, left] {\footnotesize $-\frac{1}{6}$} 
                            -- (C) node[midway, below] {\footnotesize $1$} 
                            -- cycle;

    \end{axis}
\end{tikzpicture}}
        \caption{}
        \label{fig:velocity_closure}
    \end{subfigure}
    \caption{\textbf{Kinematic closure.} (a) Dimensionless spreading time $T=t_s/\tau_{\mathrm{ic}}$ versus the unified damping parameter $\Lambda\mathcal{E}^2$. (b) Normalized characteristic spreading velocity versus the velocity damping complex $\Lambda \mathcal{E}^2 / T$. Data comprises 434 experiments from this work and \cite{liu2025maximum,aksoy2022spreading}.}
\end{figure}

Applying these boundaries eliminates the dependence on $V$, yielding a timescale:
\begin{equation}
    T \approx T_0 \left[ 1 + (\Lambda \mathcal{E}^{2})^{1/5} \right]^{-1}.
    \label{eq:time_approx}
\end{equation}
Fig.~\ref{fig:time_closure} demonstrates that this sequential closure achieves a robust collapse of the experimental data of this work, the literature \cite{aksoy2022spreading,liu2023ionic}, as well as the well-established Navier-Stokes solution for the range $\Oh = 10^{-3}$ and $\Oh = 10^{-2}$ for $10 \le \We \le 10^3$ \cite{gordillo2019theory}. 
In Supplementary Material Section S6, we show that while Lagubeau et al. \cite{lagubeau2012spreading} proposed an empirical timescale that captures their wide range of experiments, it does not extend to the broader datasets of \cite{aksoy2022spreading,liu2025maximum}, whereas the present timescale closure in Eq.~\ref{eq:time_approx} remains predictive.


To close the kinematic description (Eq.~\eqref{eq:gorin_relation}), we must couple this unified timescale with the physically relevant length and velocity scales. 

For the length scale, we distinguish the \textit{contact} diameter $D_c$ from the \textit{optical} width $D_w$. Because $D_{w}$ remains finite even as true contact vanishes ($D_{c}\rightarrow 0$) at low $\We$, utilizing $D_{c}$ provides a stricter bound on the active viscous dissipation area. This distinction is strictly necessary to determine $V_s$ and unify literature data. We derive an approximate geometric relationship, $\beta_w \approx \frac{1}{4}\beta_c^2 + 1$ for $\beta_c \leq 2$ (see Supplemental Material S1), ensuring the length scale reflects the physically active dissipation area.

Regarding spreading velocity, existing models either assume kinematic linearity ($V_s \sim V_0$) \cite{clanet2004maximal,liu2025maximum,lagubeau2012spreading}, or apply regime-specific hydrodynamic braking \cite{roisman2009inertia,gordillo2019theory,abbot2024spreading,cheng2022drop} and empirical power-law fits \cite{aksoy2022spreading}. None provide a unified physical scaling valid across the full range $10^{-3}\le \Oh \le 10^{3}$ that satisfies the energy balance. 

The energy balance (Eq.~\eqref{eq:master}) imposes a constraint on the scaling of spreading velocity as:
\begin{equation}
        V \approx \sqrt{\mathcal{E}} \left[ T^6 + \Lambda \mathcal{E}^2 T^5 \right]^{-1/6}. 
    \label{eq:velocity_closure}
\end{equation}
Because $T$ is independently approximated by Eq.~\eqref{eq:time_approx}, the energy balance reduces to an equation with a single unknown $V$. Consequently, Eq.~\eqref{eq:velocity_closure} self-consistently closes the characteristic velocity without circularity or empirical inputs.

This theoretically resolves the 'Glycerol-Water Paradox' observed in experiments \cite{abbot2024spreading,jorgensen2024deformation,aksoy2022spreading,liu2025maximum}, where viscous fluids spread faster than inviscid ones at identical $V_0$. Furthermore, at low-energy deposition ($\We \ll 1$), Eq.~\eqref{eq:velocity_closure} correctly predicts $V > 1$ due to capillary enhancement. The agreement with experiments from this work as well as from \cite{jorgensen2024deformation,aksoy2022spreading,liu2025maximum} is shown in Fig.~\ref{fig:velocity_closure}.


To achieve a unified description for the maximum spreading, we first substitute the definitions of $V$ and $T$ into Eq.~\eqref{eq:gorin_relation}, leading to $\beta_{\max} = V T \sqrt{\mathrm{We}}$. Consequently, substituting the timescale equation Eq.~\eqref{eq:time_analytical} yields:
\begin{equation}
    \beta_{\max} = \frac{\sqrt{\mathrm{We} \cdot \mathcal{E}}}{1 + (\Lambda V \mathcal{E}^{3/2})^{1/5}}.
    \label{eq:beta_V}
\end{equation}
This is dependent on $V$ and can be further expanded by substituting $V$ from the velocity equation (Eq.~\eqref{eq:velocity_closure}):
\begin{equation}
    \beta_{\max} = \frac{\sqrt{\mathrm{We} \cdot \mathcal{E}}}{1 + (\Lambda \mathcal{E}^2)^{1/5} \left[ T^6 + \Lambda \mathcal{E}^2 T^5 \right]^{-1/30}}.
    \label{eq:beta_V_expanded}
\end{equation}

Because the dynamic variation of $V$ within the viscous correction is heavily suppressed by fractional powers -1/30 (see Supplementary Material S5), it can be approximated at leading order as $V \sim \sqrt{\mathcal{E}}$, reducing the damping term to $\sim \Lambda \mathcal{E}^2$. This controlled approximation closure preserves the inertial and viscous asymptotic limits while providing a unified description across the full $(\mathrm{We},\mathrm{Re})$ space without the need for adjustable or fitted parameters. Furthermore, this leads to an approximation in a compact form that does not depend on $V$, with an error bounded by $5.2\%$ in the viscous regime (where the intermediate velocity departure is compensated by $T_0^{1/6} \approx 0.95$), and a maximum error in the intermediate regime below $9\%$. The compact explicit approximate form is:
\begin{equation}
    \beta_{\max} \approx \frac{\sqrt{\mathrm{We} \cdot \mathcal{E}}}{1 + (\Lambda \mathcal{E}^2)^{1/5}},
    \label{eq:beta_master}
\end{equation}
which is only a function of the known dimensionless numbers $\mathrm{Oh}$ and $\mathrm{We}$ and can be written as:
\begin{equation}
    \beta_{\max} \approx \frac{\sqrt{\mathrm{We} + 12}}{1 + \mathrm{Oh}^{1/5} (\mathrm{We} + 12)^{2/5}}.
    \label{eq:beta_OhWe}
\end{equation}


In the inertial limit ($\Lambda \to 0, \We \gg 1$), the denominator of Eq.~\eqref{eq:beta_master} approaches unity, recovering the classical $\beta_{\mathrm{max}} \sim \We^{1/2}$ scaling. Conversely, in the viscous limit ($\Lambda \gg 1$), the damping term dominates to yield $\beta_{\mathrm{max}} \sim \Lambda^{-1/5} \We^{1/2} \sim \Re^{1/5}$, a result consistent with both boundary-layer and lubrication theory.

\begin{figure}[t!]
    \centering
    \input{Figures/beta_master}
    \caption{\textbf{Unified scaling of maximum spreading.} Normalized spreading ratio $\beta_{\mathrm{max}} / \sqrt{\We \cdot \mathcal{E}}$ based on contact diameter $D_c$ plotted against the unified damping parameter $\Lambda \mathcal{E}^2$, following Eq.~\eqref{eq:beta_V_expanded}, \eqref{eq:beta_master}, and \eqref{eq:beta_OhWe}. The curve collapses $\sim 1,000$ experiments spanning extensive physical ranges (surface tension $0.02$--$0.5$~N/m, viscosity $10^{-3}$--$45$~Pa$\cdot$s, diameter $30~\mu\text{m}$--$15$~mm, impact velocity $\le 40$~m/s, and contact angle $\theta \in [0^\circ, 180^\circ]$), obtained in this work and sourced from the literature \cite{jorgensen2024deformation,liu2025maximum,lee2016universal,laan2014maximum,du2021analytical,liu2023ionic,roisman2009inertia,cheng1977dynamic,pasandideh1996capillary,mclauchlan2025bouncing,marmanis1996scaling,naoe2014experimental,iqbal2025droplet,abbot2025nanoparticles,jambon2020deformation,sanjay2025unifying}. Mean error is $10.15\%$.}
    \label{fig:beta_master}
\end{figure}

A quantitative assessment of the proposed unified closure yields a mean error of just $10.15\%$ for all wetting impacts ($\theta < 180^\circ$), demonstrating high predictive accuracy across physical regimes and length scales. A statistical benchmark against the Pad\'e model of Laan et al.~\cite{laan2014maximum} and the energy-balance model of Pasandideh-Fard et al.~\cite{pasandideh1996capillary} confirms that Eq.~\eqref{eq:beta_OhWe} achieves the lowest global and regime-by-regime 
root-mean-square error ($\mathrm{RMS} \le 0.21$, see Supplementary Material Section S7 and Table~S1).

Minor refinements, such as $\beta_0$ wettability corrections \cite{lee2016universal,worner2023maximum}, toroidal rim momentum sinks \cite{gordillo2019theory}, leading contact line dissipation, non-uniform film thickness, detailed boundary layer effects, internal and temporal velocity profile variations, splash losses, pressure work, internal circulation, or gravitational effects \cite{liu2025maximum, jorgensen2024deformation}, remain secondary to the primary energy balance.  Furthermore, although the asymptotic exponent $\beta \sim \Re^{1/3}$ of the deposition limit for highly viscous liquids ($\Oh > 100$) \cite{jorgensen2024deformation} is not recovered analytically from our volume-limited lubrication assumption, as it lies outside the principal asymptotic theory, the resulting kinematic curve remains within 30\% of the low-$\Re$ data of \cite{jorgensen2024deformation} over the entire range considered (see Fig.~\ref{fig:beta_master}).

The unified closure assumes actual fluid-solid contact ($\theta < 180^\circ$). It does not cover perfectly non-wetting or levitating impacts like balloons \cite{jambon2020deformation}, the Leidenfrost and inverse-Leidenfrost effects \cite{roisman2018thermal,isukwem2025inverse}, or idealized $\theta = 180^\circ$ simulations \cite{gabbard2025drop, sanjay2025unifying}. Nevertheless, for highly deformed impacts ($\beta_w \ge 2$), the optical width closely approximates the physical contact diameter ($D_w \approx D_c$), allowing even idealized non-wetting cases to align empirically with the proposed scaling (Fig.~\ref{fig:beta_master}).

In summary, rather than prescribing spreading time and velocity from asymptotic limits, we resolve the energy balance to obtain a unified description governed by $\Lambda = \We^{2} \Oh$. The resulting scaling law for the maximum spreading, free from adjustable prefactors, collapses data over wide parameter ranges and resolves deviations from prior interpolations. This framework demonstrates that drop impact is a continuous process, amenable to a single predictive theory.

\textit{Acknowledgments} — We gratefully acknowledge Ilia V. Roisman, Peichun Amy Tsai, Detlef Lohse, Vatsal Sanjay, Yunus Tansu Aksoy, Loren J\o rgensen, Lihui Liu, and Etienne Jambon-Puillet for both valuable discussions and providing experimental data. We also thank Martin Wörner and Zhifeng Hu for sharing their datasets. We thank Paul Kolpakov, Twan Spuijbroek, Eise Zimmerman, Jort Cornelissen, and Tessa Koekoek, for their help with the experimental setup and data collection.

This work is supported by the European Research Council (ERC) under the European Union’s Horizon Europe research and innovation programme (Grant agreement No. 101142159).

\bibliographystyle{apsrev4-2} 
\bibliography{references}

\end{document}


\preprint{APS/123-QED}

\title{Supplementary Material for: \\
    Kinematic Closure of Drop Impact}

\author{Mete Abbot}
 \email{m.abbot@uva.nl}
 \affiliation{Van der Waals-Zeeman Institute, Institute of Physics, University of Amsterdam, Science Park 904, 1098 XH Amsterdam, Netherlands}
\affiliation{Department of Aerospace Engineering, Faculty of Engineering, Dokuz Eyl\"ul University, 35390 Izmir, Turkey}

\author{Daniel Bonn}
 \email{d.bonn@uva.nl}
 \affiliation{Van der Waals-Zeeman Institute, Institute of Physics, University of Amsterdam, Science Park 904, 1098 XH Amsterdam, Netherlands}

\date{\today}

\maketitle


\clearpage
\newpage

\setcounter{equation}{0}
\setcounter{figure}{0}
\setcounter{table}{0}
\setcounter{section}{0}
\setcounter{page}{1}

\renewcommand{\theequation}{S\arabic{equation}}
\renewcommand{\thefigure}{S\arabic{figure}}
\renewcommand{\thetable}{S\arabic{table}}
\renewcommand{\thesection}{S\arabic{section}}

\section{S1. Geometric Unification: Contact vs. Optical Diameter}
\label{sec:geometry}

A pervasive inconsistency in the drop impact literature is the interchangeable use of the maximum optical width ($D_{w}$) and the contact diameter ($D_{c}$). While $D_{w} \approx D_{c}$ for high-deformation impacts ($\text{We} \gg 1$), these scales diverge significantly in the low-deformation deposition limit ($\text{We} \ll 1$), as shown in Fig.~\ref{fig:diameters}.

As impact energy vanishes, the contact diameter approaches zero ($D_{c} \to 0$), whereas the optical width asymptotes to the initial diameter ($D_{w} \to D_{0}$). Consequently, scaling laws based on $D_{w}$ artificially imply finite viscous dissipation area even when the wetted area vanishes, leading to predictive failure in the viscous-deposition regime. Since viscous dissipation is determined by shear at the liquid-solid interface, $D_c$ is the physically relevant length scale.

To unify the data across regimes, we derive a geometric correlation based on a spherical cap approximation. For a spherical cap (Fig.~\ref{fig:geometry_explanation}) of height $h$, contact diameter $D_c$, and curvature diameter $D_w$, the Pythagorean theorem yields:
\begin{equation}
    \left(\frac{D_w}{2}\right)^2 = \left(h - \frac{D_w}{2}\right)^2 + \left(\frac{D_c}{2}\right)^2.
\end{equation}
Defining the dimensionless ratios $\beta_w = D_w/D_0$ and $\beta_c = D_c/D_0$, and assuming the low-deformation limit ($h \approx D_0$), this simplifies to a piecewise function:
\begin{equation}
    \beta_w \approx 
    \begin{cases} 
       \frac{1}{4}\beta_c^2 + 1  & \text{for } \beta_c \le 2 \\
       \beta_c & \text{for } \beta_c > 2 
    \end{cases}
    \label{eq:geometry_corr}
\end{equation}
This correction ensures that the length scale used in the energy balance represents the physically active dissipation area. The agreement of this simple analytical approximation with a large body of experiments on various substrates is shown in Fig.~\ref{fig:c_vs_w}.

Finally, the geometric correlation between the contact angle at maximum spreading and contact diameter is known to be \cite{worner2023maximum,de2019predicting,lee2016universal}:
\begin{equation}
    \beta_c(\theta) = \sqrt[3]{\frac{4 \sin^3 \theta}{(2 + \cos \theta)(1 - \cos \theta)^2}}.
\end{equation}
However, the angle at maximum spreading state remains a subject of debate \cite{vadillo2009dynamic,worner2023maximum,de2019predicting,abbot2025nanoparticles}. Nevertheless, once the maximum spreading diameter is known, the algebraic closed-form using Cardano's method yields:
\begin{equation}
\begin{split}
    \theta(\beta_c) = 2 \arctan \Biggl[ & \sqrt[3]{\frac{4}{\beta_c^3} + \sqrt{\frac{16}{\beta_c^6} + 1}} \\
    & - \sqrt[3]{-\frac{4}{\beta_c^3} + \sqrt{\frac{16}{\beta_c^6} + 1}} \Biggr]
\end{split}
\end{equation}
This equation could be tested with experimental data in future work.

\begin{figure}[h!]
    \begin{subfigure}[c]{0.3\textwidth}
        \centering
\resizebox{\linewidth}{!}{%
\begin{tikzpicture}[font=\large, >=latex]
    \tikzset{
        droplet/.style={fill=cyan!30, draw=blue!50!black, very thick},
        substrate/.style={fill=black!80},
        dimline/.style={draw=red!80!black, <->, very thick},
        guide/.style={draw=red, dashed, very thick}
    }

    \begin{scope}[xshift=0.5cm]
        \draw[droplet] (0, 3.5) circle (0.85cm);
        \draw[->, very thick] (0, 2.3) -- (0, 1);

        \fill[substrate] (-1.7, 0) rectangle (1.7, -0.5);

        \draw[droplet] (-1.5, 0) arc (180:0:1.5cm and 0.7cm) -- cycle;

        \draw[guide] (-1.5, 0) -- (-1.5, -1.0);
        \draw[guide] (1.5, 0) -- (1.5, -1.0);
        \draw[dimline] (-1.5, -0.8) -- (1.5, -0.8) node[midway, fill=white, inner sep=2pt] {$D_w = D_c$};
    \end{scope}

    \begin{scope}[xshift=4.2cm]
        \draw[droplet] (0, 3.5) circle (0.85cm);
        \draw[->, very thick] (0, 2.3) -- (0, 1.8);

        \fill[substrate] (-1.7, 0) rectangle (1.7, -0.5);

        \begin{scope}
            \clip (-1.6, 0) rectangle (1.6, 2.5); 
            \draw[droplet] (0, 0.7) circle (0.9cm); 
        \end{scope}
        \draw[droplet] (-0.6, 0) -- (0.6, 0);

        \draw[guide] (-0.9, 0.7) -- (-0.9, -1.5);
        \draw[guide] (0.9, 0.7) -- (0.9, -1.5);
        \draw[dimline] (-0.9, 0.7) -- (0.9, 0.7) node[midway, fill=cyan!30, text=black, inner sep=1pt] {$D_w$};

        \draw[guide] (-0.6, 0) -- (-0.6, -1.0);
        \draw[guide] (0.6, 0) -- (0.6, -1.0);
        \draw[dimline] (-0.6, -0.8) -- (0.6, -0.8) node[midway, fill=white, inner sep=2pt] {$D_c$};

        \node[below] at (0, -1.1) {$D_w \neq D_c$};

    \end{scope}
\end{tikzpicture}
}
        \caption{}
        \label{fig:diameters}
    \end{subfigure}
    \begin{subfigure}[c]{0.4\textwidth}
        \centering
        \begin{tikzpicture}[scale=0.65, >=latex]

    \pgfmathsetmacro{\R}{2.5}       
    \pgfmathsetmacro{\h}{0.8}       
    \pgfmathsetmacro{\subY}{-(\R-\h)} 
    \pgfmathsetmacro{\Dc}{2*sqrt(\R^2 - (\R-\h)^2)} 
    \pgfmathsetmacro{\ContactX}{sqrt(\R^2 - (\R-\h)^2)}
    
    \draw[gray!60, dashed, thick] (0,0) circle (\R);

    \begin{scope}
        \clip (-\R-0.5, \subY) rectangle (\R+0.5, \R+0.5);
        \fill[cyan!10] (0,0) circle (\R);
        \draw[blue!70!black, very thick] (0,0) circle (\R);
    \end{scope}

    \draw[thick] (-3.5, \subY) -- (3.5, \subY);
    \fill[pattern=north east lines, pattern color=gray!50] (-3.5, \subY) rectangle (3.5, \subY-0.3);
    \node[anchor=north] at (3.2, \subY-0.3) {Solid};

    \coordinate (Center) at (0,0);
    \coordinate (ContactRight) at (\ContactX, \subY);
    \coordinate (BaseCenter) at (0, \subY);

    \draw[red, thick] (Center) -- (ContactRight) node[midway, above right, font=\footnotesize] {$D_w/2$};
    \draw[black, dashed] (Center) -- (BaseCenter);
    \draw[black, dashed] (BaseCenter) -- (ContactRight);

    \draw (0.2, \subY) -- (0.2, \subY+0.2) -- (0, \subY+0.2);

    
    \draw[<->] (0, \subY-0.15) -- (\ContactX, \subY-0.15);
    \node[anchor=north] at ({0.5*\ContactX}, \subY-0.15) {\small $D_c/2$};

    \node[anchor=east] at (0, \subY/2) {\small $D_w/2 - h$};

    \draw[<->, thin] (-3, \subY) -- (-3, \R);
    \node[fill=white, inner sep=1pt] at (-3, {(\R+\subY)/2}) {$h \approx D_0$};
    \draw[gray, dotted] (0,\R) -- (-3.2, \R); 

    \draw[<->, black] (-2.5, 2.5) -- (2.5, 2.5);
    \node[fill=white, text=black] at (0, 2.5) {$D_w \approx D_0$};


\end{tikzpicture}
        \caption{}
        \label{fig:geometry_explanation}
    \end{subfigure}
        \begin{subfigure}[c]{0.49\textwidth}
        \centering
        \pgfplotsset{compat=1.17} 

\definecolor{mycolor0}{RGB}{50,155,105}    
\definecolor{mycolor5}{RGB}{213,81,156}    
\definecolor{mycolor10}{RGB}{54,80,163}    
\definecolor{mycolor125}{RGB}{244,196,48}  
\definecolor{mycolor15}{RGB}{245,127,34}   
\definecolor{mycolorSplash}{RGB}{204,204,255} 
\definecolor{mycolorNo}{RGB}{224,224,224}     

\begin{tikzpicture}
\begin{axis}[
width=1\figW,
height=1\figH,
at={(0\figW,0\figH)},
    xlabel={\footnotesize Contact Ratio $\beta_c$},
    ylabel={\footnotesize Width Ratio $\beta_w$},
    xmin=0, xmax=5,
    ymin=0, ymax=6,
    minor x tick num=1,  
    minor y tick num=1,  
    legend style={at={(0.03,0.97)}, anchor=north west, draw=none, font=\scriptsize},
    ticklabel style = {font=\scriptsize}, legend cell align=left
    ]

\addplot[dashed, color=black, line width=2.0pt]
table[row sep=crcr]{
0	1	\\
0.2	1.01	\\
0.4	1.04	\\
0.6	1.09	\\
0.8	1.16	\\
1	1.25	\\
1.2	1.36	\\
1.4	1.49	\\
1.6	1.64	\\
1.8	1.81	\\
2	2	\\
3 3 \\
4 4 \\
5 5 \\
};
\addlegendentry{Eq.~\ref{eq:geometry_corr}}

\addplot[only marks, color=blue, mark=o, mark size=1.5pt, line width=0.5pt]
table[row sep=crcr]{
0.85152	1.1531	\\
1.11514	1.3244	\\
1.17301	1.34342	\\
1.21945	1.35993	\\
1.30497	1.45796	\\
1.33906	1.47159	\\
1.19672	1.47232	\\
1.33278	1.47819	\\
1.40727	1.51661	\\
1.27037	1.5428	\\
1.46236	1.56285	\\
1.4547	1.58595	\\
1.46904	1.59407	\\
1.54012	1.60923	\\
1.5538	1.63166	\\
1.56679	1.68111	\\
1.62938	1.68346	\\
1.63472	1.74448	\\
1.71158	1.75351	\\
1.66023	1.75849	\\
1.5274	1.7597	\\
1.72084	1.776	\\
1.71587	1.79852	\\
1.7891	1.83141	\\
1.79208	1.84712	\\
1.7769	1.84946	\\
1.75245	1.8501	\\
1.80845	1.8788	\\
1.8577	1.90692	\\
1.7131	1.91541	\\
1.89755	1.93926	\\
1.84778	1.93986	\\
1.89305	1.94839	\\
1.96829	2.02919	\\
1.96538	2.05042	\\
2.02308	2.06955	\\
1.89194	2.08867	\\
2.06538	2.102	\\
2.04885	2.11076	\\
2.04155	2.11916	\\
2.11632	2.19402	\\
2.05518	2.24487	\\
2.18698	2.25791	\\
2.34153	2.39723	\\
2.2505	2.42611	\\
2.37775	2.43675	\\
2.44701	2.5789	\\
2.57327	2.6899	\\
2.69434	2.85514	\\
2.71887	2.87059	\\
};
\addlegendentry{\cite{hu2025reconsideration}}

\addplot[only marks, color=red, mark=square, mark size=1.5pt, line width=0.5pt]
table[row sep=crcr]{
0.205645161	1.008064516	\\
0.216718266	1.00619195	\\
0.218253968	1	\\
0.220820189	1.001577287	\\
0.221774194	1	\\
0.227074236	1.008733624	\\
0.230046948	1.004694836	\\
0.234817814	1	\\
0.234817814	1.004048583	\\
0.235294118	1.009049774	\\
0.240566038	1.004716981	\\
0.24291498	1.004048583	\\
0.247058824	1	\\
0.25	1	\\
0.290697674	1.023255814	\\
0.293103448	1	\\
0.294117647	1.011764706	\\
0.3	1.005882353	\\
0.310204082	1.012244898	\\
0.310734463	1.02259887	\\
0.3125	1.009615385	\\
0.313953488	1.023255814	\\
0.313953488	1.01744186	\\
0.313953488	1.01744186	\\
0.315789474	1.010526316	\\
0.316326531	1	\\
0.317948718	1.005128205	\\
0.318181818	1.01010101	\\
0.320512821	1.008547009	\\
0.323943662	1.009389671	\\
0.324468085	1.005319149	\\
0.325581395	1.011627907	\\
0.325714286	1.011428571	\\
0.325966851	1.005524862	\\
0.326315789	1.005263158	\\
0.326315789	1.005263158	\\
0.327102804	1.014018692	\\
0.327433628	1.013274336	\\
0.333333333	1.005649718	\\
0.333333333	1.005649718	\\
0.336734694	1.005102041	\\
0.3374613	1.00619195	\\
0.337606838	1.008547009	\\
0.338383838	1.01010101	\\
0.342857143	1.011428571	\\
0.343023256	1.011627907	\\
0.347058824	1.011764706	\\
0.358255452	1.012461059	\\
0.358255452	1.00623053	\\
0.358870968	1.008064516	\\
0.362745098	1.013071895	\\
0.366935484	1.016129032	\\
0.373015873	1.003968254	\\
0.373088685	1.012232416	\\
0.375384615	1.012307692	\\
0.380062305	1.009345794	\\
0.384615385	1.01183432	\\
0.384615385	1.014423077	\\
0.384615385	1.004524887	\\
0.3894081	1.009345794	\\
0.389830508	1.005649718	\\
0.392344498	1.009569378	\\
0.394117647	1.005882353	\\
0.395721925	1.010695187	\\
0.396551724	1.011494253	\\
0.396825397	1.00952381	\\
0.397196262	1.009345794	\\
0.4	1.01025641	\\
0.401826484	1.00913242	\\
0.40397351	1.016556291	\\
0.406349206	1.00952381	\\
0.406349206	1.00952381	\\
0.408945687	1.015974441	\\
0.412903226	1.016129032	\\
0.422222222	1.019047619	\\
0.422222222	1.019047619	\\
0.444444444	1.015151515	\\
0.445595855	1.020725389	\\
0.446808511	1.008510638	\\
0.452631579	1.031578947	\\
0.452631579	1.015789474	\\
0.453608247	1.020618557	\\
0.455284553	1.012195122	\\
0.456521739	1.008695652	\\
0.457777778	1.008888889	\\
0.463203463	1.012987013	\\
0.473958333	1.026041667	\\
0.525993884	1.027522936	\\
0.529230769	1.033846154	\\
0.532110092	1.039755352	\\
0.600682594	1.040955631	\\
0.605015674	1.053291536	\\
0.608562691	1.064220183	\\
0.628865979	1.067010309	\\
0.630208333	1.052083333	\\
0.644670051	1.07106599	\\
0.646153846	1.076923077	\\
0.663316583	1.075376884	\\
0.668380463	1.066838046	\\
0.711409396	1.090604027	\\
0.726666667	1.103333333	\\
0.867507886	1.164037855	\\
0.878136201	1.168458781	\\
0.887096774	1.164516129	\\
0.888501742	1.177700348	\\
0.894736842	1.153110048	\\
0.896226415	1.155660377	\\
0.897196262	1.14953271	\\
0.902777778	1.162037037	\\
0.902777778	1.162037037	\\
0.908653846	1.149038462	\\
0.919811321	1.16509434	\\
0.93989071	1.174863388	\\
0.946859903	1.183574879	\\
0.951612903	1.188172043	\\
0.956284153	1.191256831	\\
0.962365591	1.188172043	\\
0.968253968	1.195767196	\\
0.995475113	1.199095023	\\
1.008928571	1.21875	\\
1.022026432	1.233480176	\\
1.035714286	1.253246753	\\
1.046413502	1.244725738	\\
1.054187192	1.275862069	\\
1.067741935	1.270967742	\\
1.085987261	1.302547771	\\
1.088328076	1.29022082	\\
1.100946372	1.29022082	\\
1.105990783	1.276497696	\\
1.111627907	1.288372093	\\
1.122065728	1.300469484	\\
1.130232558	1.288372093	\\
1.141141141	1.342342342	\\
1.146268657	1.334328358	\\
1.156682028	1.290322581	\\
1.158490566	1.328301887	\\
1.158878505	1.339563863	\\
1.164179104	1.340298507	\\
1.167286245	1.334572491	\\
1.167286245	1.319702602	\\
1.170149254	1.340298507	\\
1.171428571	1.371428571	\\
1.174041298	1.345132743	\\
1.18079096	1.38700565	\\
1.183800623	1.358255452	\\
1.187725632	1.335740072	\\
1.197492163	1.36677116	\\
1.203571429	1.335714286	\\
1.205047319	1.38170347	\\
1.219512195	1.300813008	\\
1.221183801	1.38317757	\\
1.227436823	1.364620939	\\
1.234432234	1.384615385	\\
1.234636872	1.410614525	\\
1.241877256	1.389891697	\\
1.257142857	1.389285714	\\
1.272727273	1.428152493	\\
1.288629738	1.425655977	\\
1.300291545	1.425655977	\\
1.309192201	1.440111421	\\
1.310541311	1.444444444	\\
1.331683168	1.462871287	\\
1.341584158	1.472772277	\\
1.35	1.4675	\\
1.361386139	1.48019802	\\
1.361386139	1.472772277	\\
1.3625	1.4875	\\
1.37254902	1.50245098	\\
1.381995134	1.501216545	\\
1.461538462	1.461538462	\\
1.722891566	1.722891566	\\
2.061946903	2.061946903	\\
2.294536817	2.294536817	\\
2.33640553	2.33640553	\\
2.394209354	2.420935412	\\
2.543859649	2.587719298	\\
2.585034014	2.641723356	\\
2.594654788	2.650334076	\\
2.605790646	2.661469933	\\
2.731070496	2.731070496	\\
3.359281437	3.359281437	\\
3.473988439	3.473988439	\\
3.607427056	3.607427056	\\
};
\addlegendentry{\cite{jorgensen2024deformation}}

\end{axis}
\end{tikzpicture}
        \caption{}
        \label{fig:c_vs_w}
    \end{subfigure}
    \caption{(a) Two cases in which the contact diameter ($D_c$) is (left) and is not (right) the same as the maximum width of a droplet ($D_w$). (b) Spherical cap model for the low deformation limit used to derive Eq.~\eqref{eq:geometry_corr}. (c) Agreement of the geometrical relationship between $D_c$ and $D_w$ (dashed line, Eq.~\eqref{eq:geometry_corr}) with 231 experimental data points from \cite{jorgensen2024deformation,hu2025reconsideration} representing various impact conditions.}
\end{figure}


\section{S2. Formulation of the Master Energy Balance}
\label{sec:energy_formulation}

We derive the governing equation by conserving energy from the moment of impact ($t=0$) to the moment of maximum spreading ($t=t_s$). The total initial energy (kinetic $E_k$ plus surface $E_{s,0}$) is balanced against the final surface energy ($E_{s,f}$) and the total viscous dissipation ($W_{\nu}$):
\begin{equation}
    E_k + E_{s,0} \approx E_{s,f} + W_{\nu}.
    \label{eq:S_balance_gen}
\end{equation}
To obtain the dimensionless master equation (Eq. 2 in the main text), we normalize each term by the initial kinetic energy $E_k \sim \rho D_0^3 V_0^2$. The left side of the balance becomes the total dimensionless energy density $\mathcal{E} = 1 + 12/\We$, where $12/\We$ represents the ratio of initial spherical surface energy to kinetic energy.

\subsection*{A. Final Surface Energy $(E_{s,f})$}

While regime-specific models often justifiably neglect surface energy variations \cite{liu2025maximum,worner2023maximum}, a kinematic closure must retain capillary storage to capture the full parameter space. 

The exact change in surface energy is given by:
\begin{equation}
    \Delta E_{\sigma} = \sigma (A_{LV} - A_0) - \sigma A_{SL} \cos \theta_{\mathrm{dyn}}
    \label{eq:deltaasigma}
\end{equation}
where $\theta_{\mathrm{dyn}}$ is the dynamic contact angle. Using the spherical cap model of \cite{worner2023maximum} for surface area leads to the change of surface area as:
\begin{equation}
\begin{split}
    A_{LV} / A_0 &= \frac{\sqrt[3]{\beta^6 + 8\sqrt{\beta^6 + 16} + 32} - \beta^2}{4} \\
    &\quad + \frac{\beta^4}{4\sqrt[3]{\beta^6 + 8\sqrt{\beta^6 + 16} + 32}}
    \label{eq:hardmodel}
\end{split}
\end{equation}
The equation is presented in Fig.~\ref{fig:spherical_cap} as compared to a disk and to a cylinder approximation. Substituting Eq.~\ref{eq:hardmodel} into Eq.~\ref{eq:deltaasigma} gives the total surface energy change for a perfect spherical cap.

Given that drops do not make perfect spherical caps, a simplified spherical cap interpolation model of Eq.~\ref{eq:hardmodel} was presented by \cite{worner2023maximum}, which leads to the change in surface energy as:
\begin{equation}
    \Delta E_{\sigma} \approx \frac{\pi}{4} \sigma D_0^2 \beta^2 \left[ 3 \left( \frac{\beta^2}{\beta^2 + 4} - \cos \theta_{\mathrm{dyn}} \right) \right].
\end{equation}

However, even with this simplification, additional complexities arise. Firstly, the spherical cap model is too idealized and does not represent the real drop shape with possible rim formation. Secondly, the precise dynamic angle to be used during impact remains heavily debated \cite{vadillo2009dynamic, de2019predicting, iqbal2025droplet, abbot2025nanoparticles}. In the recent work of \cite{quetzeri2019role}, it was shown that the relevant dynamic angle is $\theta_{\mathrm{max}}$, defined as the asymptotic maximum advancing dynamic contact angle achieved during the rapid spreading phase of the impacting droplet. This angle remains tightly bounded between $87^{\circ}$ and $150^{\circ}$ for spreading drops, regardless of the liquid-substrate combination (from fully wetting to superhydrophobic). Consequently, the contact angle penalty $\cos \theta_{\mathrm{max}}$ is strictly bounded between approximately $0$ and $-0.86$. It must be noted that this angle can only be measured using drop impact experiments, which means it does not have any predictive power. 

Given all the physical limitations, we conclude that:

\begin{itemize}
    \item \textbf{Low deformation limit ($\beta \le 2$):} The absolute change in the total liquid-gas surface area is intrinsically negligible. As demonstrated by the accurate spherical cap model in \cite{worner2023maximum} (Fig.~\ref{fig:spherical_cap}), $\Delta E_\sigma \approx 0$ for $\beta \le 2$, rendering the wettability term irrelevant.
    \item \textbf{High-energy impacts ($\We \gg 1$):} The ratio of surface energy change to kinetic energy scales as $\frac{\Delta E_\sigma}{E_k} \sim \frac{1}{\We}(\beta^2 - 1)$. Consequently, the entire surface energy penalty vanishes relative to inertia ($\frac{\Delta E_\sigma}{E_k} \rightarrow 0$) for high $\We$ impacts \cite{roisman2009inertia, liu2025maximum}.
    \item \textbf{High deformations ($\beta \gg 1$):} When capillary forces are relevant at large spreading ratios, the generation of new liquid-gas interfacial area ($\sim \beta^2$) strictly dominates the bounded $\cos \theta$ contribution.
\end{itemize}

\begin{figure}[h!]
    \centering
    \pgfplotsset{compat=1.17} 

\definecolor{mycolor0}{RGB}{50,155,105}    
\definecolor{mycolor5}{RGB}{213,81,156}    
\definecolor{mycolor10}{RGB}{54,80,163}    
\definecolor{mycolor125}{RGB}{244,196,48}  
\definecolor{mycolor15}{RGB}{245,127,34}   
\definecolor{mycolorSplash}{RGB}{204,204,255} 
\definecolor{mycolorNo}{RGB}{224,224,224}     

\begin{tikzpicture}
\begin{axis}[
width=0.9\figW,
height=0.8\figH,
at={(0\figW,0\figH)},
    xlabel={\footnotesize $\beta_c$},
    ylabel={\footnotesize Normalized Surface Area},
    xmin=0, xmax=5,
    ymin=0, ymax=8,
    ytick={0, 2, 4, 6},
    minor x tick num=1,  
    minor y tick num=1,  
    legend style={at={(0.01,0.99)}, anchor=north west, draw=none, font=\scriptsize},
    ticklabel style = {font=\scriptsize}, legend cell align=left
    ]

\addplot[color=black, line width=2.0pt]
table[row sep=crcr]{
0	1	\\
0.1	0.99750626	\\
0.2	0.990100667	\\
0.3	0.97801384	\\
0.4	0.961642595	\\
0.5	0.941568329	\\
0.6	0.918581656	\\
0.7	0.893710813	\\
0.8	0.868248342	\\
0.9	0.843765748	\\
1	0.822099606	\\
1.1	0.805287645	\\
1.2	0.795436298	\\
1.3	0.794520161	\\
1.4	0.804150111	\\
1.5	0.8253819	\\
1.6	0.858637915	\\
1.7	0.903767499	\\
1.8	0.960207226	\\
1.9	1.027169843	\\
2	1.103803403	\\
2.1	1.189296164	\\
2.2	1.282930449	\\
2.3	1.38410044	\\
2.4	1.492309222	\\
2.5	1.607156209	\\
2.6	1.728321592	\\
2.7	1.855551226	\\
2.8	1.988643392	\\
2.9	2.1274378	\\
3	2.271806747	\\
3.1	2.421648122	\\
3.2	2.576879927	\\
3.3	2.737436016	\\
3.4	2.903262766	\\
3.5	3.074316501	\\
3.6	3.250561466	\\
3.7	3.431968256	\\
3.8	3.618512569	\\
3.9	3.810174232	\\
4	4.006936423	\\
4.1	4.208785056	\\
4.2	4.415708285	\\
4.3	4.627696109	\\
4.4	4.844740054	\\
4.5	5.06683291	\\
4.6	5.293968524	\\
4.7	5.526141624	\\
4.8	5.763347683	\\
4.9	6.005582794	\\
5	6.252843582	\\
5.1	6.505127119	\\
5.2	6.76243086	\\
5.3	7.024752585	\\
5.4	7.292090352	\\
5.5	7.564442461	\\
5.6	7.841807418	\\
5.7	8.124183907	\\
5.8	8.411570766	\\
5.9	8.703966968	\\
6	9.001371603	\\
};
\addlegendentry{Spherical Cap}

\addplot[dotted, color=blue, line width=2.0pt]
table[row sep=crcr]{
0.3	2.244722222	\\
0.4	1.706666667	\\
0.5	1.395833333	\\
0.6	1.201111111	\\
0.7	1.074880952	\\
0.8	0.993333333	\\
0.9	0.943240741	\\
1	0.916666667	\\
1.1	0.908560606	\\
1.2	0.915555556	\\
1.3	0.935320513	\\
1.4	0.966190476	\\
1.5	1.006944444	\\
1.6	1.056666667	\\
1.7	1.114656863	\\
1.8	1.18037037	\\
1.9	1.253377193	\\
2	1.333333333	\\
2.1	1.419960317	\\
2.2	1.513030303	\\
2.3	1.612355072	\\
2.4	1.717777778	\\
2.5	1.829166667	\\
2.6	1.946410256	\\
2.7	2.06941358	\\
2.8	2.198095238	\\
2.9	2.332385057	\\
3	2.472222222	\\
3.1	2.617553763	\\
3.2	2.768333333	\\
3.3	2.924520202	\\
3.4	3.086078431	\\
3.5	3.25297619	\\
3.6	3.425185185	\\
3.7	3.60268018	\\
3.8	3.785438596	\\
3.9	3.973440171	\\
4	4.166666667	\\
4.1	4.365101626	\\
4.2	4.568730159	\\
4.3	4.77753876	\\
4.4	4.991515152	\\
4.5	5.210648148	\\
4.6	5.434927536	\\
4.7	5.664343972	\\
4.8	5.898888889	\\
4.9	6.138554422	\\
5	6.383333333	\\
5.1	6.633218954	\\
5.2	6.888205128	\\
5.3	7.148286164	\\
5.4	7.41345679	\\
5.5	7.683712121	\\
5.6	7.959047619	\\
5.7	8.239459064	\\
5.8	8.524942529	\\
5.9	8.81549435	\\
6	9.111111111	\\
};
\addlegendentry{Cylinder}

\addplot[dotted, color=red, line width=2.0pt]
table[row sep=crcr]{
0	0	\\
0.1	0.0025	\\
0.2	0.01	\\
0.3	0.0225	\\
0.4	0.04	\\
0.5	0.0625	\\
0.6	0.09	\\
0.7	0.1225	\\
0.8	0.16	\\
0.9	0.2025	\\
1	0.25	\\
1.1	0.3025	\\
1.2	0.36	\\
1.3	0.4225	\\
1.4	0.49	\\
1.5	0.5625	\\
1.6	0.64	\\
1.7	0.7225	\\
1.8	0.81	\\
1.9	0.9025	\\
2	1	\\
2.1	1.1025	\\
2.2	1.21	\\
2.3	1.3225	\\
2.4	1.44	\\
2.5	1.5625	\\
2.6	1.69	\\
2.7	1.8225	\\
2.8	1.96	\\
2.9	2.1025	\\
3	2.25	\\
3.1	2.4025	\\
3.2	2.56	\\
3.3	2.7225	\\
3.4	2.89	\\
3.5	3.0625	\\
3.6	3.24	\\
3.7	3.4225	\\
3.8	3.61	\\
3.9	3.8025	\\
4	4	\\
4.1	4.2025	\\
4.2	4.41	\\
4.3	4.6225	\\
4.4	4.84	\\
4.5	5.0625	\\
4.6	5.29	\\
4.7	5.5225	\\
4.8	5.76	\\
4.9	6.0025	\\
5	6.25	\\
5.1	6.5025	\\
5.2	6.76	\\
5.3	7.0225	\\
5.4	7.29	\\
5.5	7.5625	\\
5.6	7.84	\\
5.7	8.1225	\\
5.8	8.41	\\
5.9	8.7025	\\
6	9	\\
};
\addlegendentry{Disk}

\addplot[dashed]
table[row sep=crcr]{
0 1 \\
6 1 \\
};

\end{axis}
\end{tikzpicture}
    \caption{Normalized liquid-gas surface area as a function of spreading ratio as obtained from the complete spherical cap model of \cite{worner2023maximum} (black line).}
    \label{fig:spherical_cap}
\end{figure}

Therefore, evaluating the geometrical prefactor of the final surface energy term as unity (yielding $E_{s,f} \sim \sigma D_{\mathrm{max}}^2$) is a leading-order thermodynamic approximation that ensures the master energy balance correctly captures the relevant change in surface energy.

To express this final surface energy in dimensionless form in the master energy balance, we apply the kinematic relation $D_{\mathrm{max}} = V_s t_s$. The final surface energy then scales as:
\begin{equation}
    E_{s,f} \sim \sigma (V_s t_s)^2.
\end{equation}
To non-dimensionalize this term, we normalize it by the initial kinetic energy $E_k \sim \rho D_0^3 V_0^2$:
\begin{equation}
    \frac{E_{s,f}}{E_k} \sim \frac{\sigma V_s^2 t_s^2}{\rho D_0^3 V_0^2}.
\end{equation}
We express the kinematics in terms of our dimensionless variables $V = V_s/V_0$ and $T = t_s/\tau_{ic}$, where $\tau_{ic} = \sqrt{\rho D_0^3 / \sigma}$ is the inertio-capillary timescale. Substituting $V_s = V V_0$ and expanding $t_s^2 = T^2 (\rho D_0^3 / \sigma)$ explicitly yields:
\begin{equation}
    \frac{E_{s,f}}{E_k} \sim \frac{\sigma (V^2 V_0^2) \left( T^2 \frac{\rho D_0^3}{\sigma} \right)}{\rho D_0^3 V_0^2}.
\end{equation}
Simplifying the physical constants strictly isolates the dimensionless variables:
\begin{equation}
    \frac{E_{s,f}}{E_k} \sim V^2 T^2.
    \label{eq:surfaceterm}
\end{equation}
This dimensionless capillary storage term naturally forms the first component on the right-hand side of the master energy equation.

\subsection*{B. Viscous Dissipation ($W_\nu$)}

To evaluate the viscous work $W_{\nu}$, we model the dissipation through the lens of volume-limited lubrication flow. 

Following phase-resolved dissipation frameworks \cite{roisman2009inertia,eggers2010drop,liu2025maximum,sanjay2025unifying}, the total viscous work can be partitioned into an early boundary-layer growth phase and a subsequent bulk lubrication phase:
\begin{equation}
    W_\nu(t_s) = \int_0^{t^*} \dot{W}_{\mathrm{BL}}(t) \, dt + \int_{t^*}^{t_s} \dot{W}_{\mathrm{lub}}(t) \, dt,
    \label{eq:phase_resolved}
\end{equation}
where $t^*$ is the crossover time at which the boundary layer thickness $\delta(t) \sim \sqrt{\nu t}$ reaches the lamella film thickness $h(t) \sim D_0^3/D(t)^2$. In the inertio-viscous regime ($\Lambda\mathcal{E}^2 \gg 1$), the boundary layer spans the thin film rapidly ($t^* \ll t_s$), so the total dissipation is overwhelmingly dominated by the second (lubrication) integral. In the inertio-capillary regime ($\Lambda\mathcal{E}^2 \ll 1$), total viscous dissipation is energetically negligible compared to surface energy ($W_\nu \ll E_\sigma$). Consequently, evaluating the volume-limited lubrication rate over the full impact duration serves as the leading-order physical reduction approximation across the majority of the parameter domain:
\begin{equation}
    \dot{W} \sim \eta (V_s/h)^2 D^2 h.
\end{equation} 
By invoking volume conservation ($V_{\text{drop}} \sim D_0^3 \sim D^2 h$), the film thickness scales as $h \sim D_0^3/D^2$. Substituting this into the dissipation rate yields:
\begin{equation}
    \dot{W} \sim \eta \frac{V_s^2}{h} D^2 \sim \eta \frac{V_s^2 D^4}{D_0^3}.
\end{equation}
By substituting the kinematic relation $D(t) \sim V_s t$, the instantaneous dissipation rate scales as $\dot{W}(t) \sim \eta \frac{V_s^6 t^4}{D_0^3}$. Integrating this rate over the total spreading time $t_s$ yields the total viscous work:
\begin{equation}
    W_\nu \sim \int_0^{t_s} \dot{W}(t) \, dt \sim \int_0^{t_s} \eta \frac{V_s^6 t^4}{D_0^3} \, dt \sim \eta \frac{V_s^6 t_s^5}{D_0^3}.
\end{equation}
To non-dimensionalize the viscous energy term, we normalize by the initial kinetic energy $E_k \sim \rho D_0^3 V_0^2$:
\begin{equation}
    \frac{W_{\nu}}{E_k} \sim \frac{\eta V_s^6 t_s^5}{\rho D_0^6 V_0^2}.
\end{equation}
Finally, we express the kinematics in terms of our dimensionless variables $V = V_s/V_0$ and $T = t_s/\tau_{ic}$, where $\tau_{ic} = \sqrt{\rho D_0^3 / \sigma}$ is the inertio-capillary timescale. Substituting $V_s = V V_0$ and expanding $t_s^5 = T^5 (\rho D_0^3 / \sigma)^{5/2}$ into the normalized viscous penalty explicitly yields:
\begin{equation}
    \frac{W_{\nu}}{E_k} \sim \frac{\eta (V^6 V_0^6) \left( T^5 \frac{\rho^{5/2} D_0^{15/2}}{\sigma^{5/2}} \right)}{\rho D_0^6 V_0^2}.
\end{equation}
Simplifying the physical constants isolates the dimensionless groups:
\begin{equation}
    \frac{W_{\nu}}{E_k} \sim \left( \frac{\eta \rho^{3/2} D_0^{3/2} V_0^4}{\sigma^{5/2}} \right) V^6 T^5.
\end{equation}
Recognizing that the leading dimensionless coefficient is exactly the product of $\We^2$ and $\Oh$, the normalized viscous penalty assumes the final form:
\begin{equation}
    \frac{W_{\nu}}{E_k} \sim \Lambda V^6 T^5,
    \label{eq:viscousterm}
\end{equation}
where $\Lambda \equiv \We^{2}\Oh$ represents the unified damping parameter presented in the main text.

\section{S3. Analytical Derivation of the Unified Timescale}
\label{sec:time_deriv}

Substituting the scaled surface energy term Eq.~\ref{eq:surfaceterm} and viscous dissipation term Eq.~\ref{eq:viscousterm} into the energy balance Eq.~\eqref{eq:S_balance_gen} yields the dimensionless energy balance equation:
\begin{equation}
    \mathcal{E} \approx V^2 T^2 + \Lambda V^6 T^5,
    \label{eq:S_poly}
\end{equation}
where $\mathcal{E} = 1 + 12/\text{We}$ is the total energy density, $V = V_s/V_0$, $T = t_s/\tau_{ic}$, and $\Lambda = \We^2 \Oh$. We solve this polynomial by matching its asymptotic limits.

\textbf{Inertial Limit ($\Lambda \to 0$):} The viscous term vanishes.
\begin{equation}
    \mathcal{E} \approx V^2 T^2 \implies T_{\mathrm{in}} \approx \frac{\sqrt{\mathcal{E}}}{V}.
\end{equation}

\textbf{Viscous Limit ($\Lambda \gg 1$):} The viscous term dominates.
\begin{equation}
    \mathcal{E} \approx \Lambda V^6 T^5 \implies T_{\mathrm{visc}} \approx \left( \frac{\mathcal{E}}{\Lambda V^6} \right)^{1/5}.
\end{equation}

\textbf{Unified Solution:} We construct a matched asymptotic interpolation using the harmonic mean of squares ($T^{-1} \approx T_{\mathrm{in}}^{-1} + T_{\mathrm{visc}}^{-1}$) to ensure smooth decay:
\begin{equation}
    T \approx \left[ \frac{V}{\mathcal{E}^{1/2}} + \frac{(\Lambda V^6)^{1/5}}{\mathcal{E}^{1/5}} \right]^{-1}.
    \label{eq:time_analytical0}
    \end{equation}
Factoring $\sqrt{\mathcal{E}}/V$, this can be rewritten as:
\begin{equation}
    T \approx \frac{\sqrt{\mathcal{E}}}{V} \left[ 1 + (\Lambda V \mathcal{E}^{3/2})^{1/5} \right]^{-1}.
    \label{eq:time_analytical}
    \end{equation}

To formulate a predictive timescale strictly decoupled from the spreading velocity, we evaluate the terms in Eq.~\eqref{eq:time_analytical} sequentially. First, the leading prefactor $\sqrt{\mathcal{E}}/V$ corresponds to the inertial limit. In this inviscid regime, the spreading time is independently governed by the local dynamic balance between the droplet's inertial force and the linear capillary restoring force as detailed in \cite{bartolo2005retraction}. 

Substituting the droplet mass $m = \frac{\pi}{6}\rho D_0^3$, the inertial deceleration force scales as:
\begin{equation}
    F_i \sim m (V_s/t_s) \sim \frac{\pi}{6}\rho D_0^3 (V_s/t_s).
\end{equation}
The capillary restoring force acting against this deformation scales as $F_c \sim \sigma D_{\mathrm{max}}$. Applying our kinematic relation $D_{\mathrm{max}} = V_s t_s$, the capillary force becomes $F_c \sim \sigma V_s t_s$. Equating these two forces ($F_i \approx F_c$) yields:
\begin{equation}
    \frac{\pi}{6} \rho D_0^3 \frac{V_s}{t_s} \approx \sigma V_s t_s.
\end{equation}
The characteristic velocity $V_s$ strictly cancels out from both sides of the balance. This physically isolates the timescale from the spreading velocity at the inviscid limit. Solving for $t_s$ gives the inertial time limit $t_s \approx \sqrt{\frac{\pi}{6}} \sqrt{\frac{\rho D_0^3}{\sigma}}$. 

Normalizing this by the inertio-capillary timescale $\tau_{ic} = \sqrt{\rho D_0^3 / \sigma}$ yields the constant geometric coefficient $T_0 = \sqrt{\pi/6}$. 

To remove the velocity dependence from the second term in the balance, substituting the constant $T_0$ into the viscous damping term of the timescale interpolation yields $(\Lambda \mathcal{E}^2 / T_0)^{1/5} \approx (\Lambda \mathcal{E}^{2})^{1/5}$, since the geometric prefactor $T_0^{-1/5} \approx 1.06 \approx 1$. 

Further mathematical justification on why the velocity in the second term can be approximated at leading order ($V = \sqrt{\mathcal{E}}$) is provided in Section S5.
    
Applying these boundaries eliminates dependence on $V$, yielding a timescale:
\begin{equation}
    T \approx T_0 \left[ 1 + (\Lambda \mathcal{E}^{2})^{1/5} \right]^{-1}.
    \label{eq:time_approx}
\end{equation}

It must be noted that while the energy balance strictly yields $(T/T_{in})^2 + (T/T_{visc})^5 \approx 1$, this fifth-order polynomial lacks a general algebraic solution. The harmonic mean utilized here is therefore not an arbitrary assumption, but a standard asymptotic matching technique (a linear superposition of dissipation rates) required to achieve a closed-form predictive scaling law. As shown in Fig.~\ref{fig:exact_vs_approx}, the harmonic approximation and the exact numerical root share identical inertial and viscous asymptotes. The harmonic mean introduces a maximum mathematical deviation of approximately 30\% exclusively in the low $\Lambda$ regime ($\Lambda \approx 0.1$). This variance is physically acceptable as it falls well within the inherent experimental scatter for low velocity impact, as these experiments are dominated by noise from drop generation methods due to being close to the needle or nozzle, such as wobbling and non-spherical drop shapes.

\begin{figure}[h]
    \centering
    \begin{tikzpicture}
        \begin{axis}[
            width=1.2\figW,
            height=1.0\figH,
            xmode=log,
            ymode=log,
            xmin=1e-5, xmax=1e10,
            ymin=0.003, ymax=2,
            xlabel={$\Lambda$},
            ylabel={$T/T_{in}$},
            legend pos=south west,
            legend style={at={(0.02,0.30)}, anchor=north west, draw=none, fill=none, font=\footnotesize},
            ticklabel style = {font=\footnotesize}, legend cell align=left
        ]
        
        \addplot [
            domain=0.0005:0.999999, 
            samples=100, 
            thick, 
            black
        ] ({ (1 - x^2)/(x^5) }, {x});
        \addlegendentry{$(T/T_{in})^2 + (T/T_{vis})^5 \approx 1$}

        \addplot [
            domain=1e-5:1e10, 
            samples=100, 
            thick, 
            red, 
            dashed
        ] {1 / (1 + x^(0.2))};
        \addlegendentry{$(T/T_{in}) + (T/T_{vis}) \approx 1$}

        \end{axis}
    \end{tikzpicture}
    \caption{Comparison of the exact numerical root of the master energy polynomial against the closed-form analytical approximation. Both curves share identical inertial and viscous asymptotes. The maximum deviation occurs in the low $\Lambda$ regime, well within the inherent scatter of experimental impact data at very low impact velocity.}
    \label{fig:exact_vs_approx}
\end{figure}

\section{S4. Analytical Derivation of the Spreading Velocity}
\label{sec:velocity_closure}

We return to the dimensionless master polynomial (Eq.~\eqref{eq:S_poly}):
\begin{equation}
    V^2 T^2 + \Lambda V^6 T^5 \approx \mathcal{E}.
\end{equation}
Similar to the timescale derivation, we solve for the velocity $V$ by first extracting its asymptotic limits. In the inviscid limit ($\Lambda \to 0$), the velocity scales as $V_{\mathrm{in}} \approx \sqrt{\mathcal{E}}/T$. In the extreme viscous limit ($\Lambda \gg 1$), the velocity scales as $V_{\mathrm{visc}} \approx (\mathcal{E} / \Lambda T^5)^{1/6}$.

Rearranging the master polynomial yields the implicit algebraic balance $(V/V_{\mathrm{in}})^2 + (V/V_{\mathrm{visc}})^6 \approx 1$. To extract a continuous explicit function across all regimes, we apply a matched asymptotic interpolation. 

To maximize numerical accuracy (as demonstrated in Fig.~\ref{fig:velocity_exact_vs_approx}), algebraically rationalize the root, and strictly preserve the integer unified parameter $\Lambda$ for subsequent kinematic reductions (calculating $\beta_{\mathrm{max}}$), we apply a 6th-power harmonic mean ($V^{-6} \approx V_{\mathrm{in}}^{-6} + V_{\mathrm{visc}}^{-6}$):
\begin{equation}
    V \approx \left[ \left( \frac{T}{\sqrt{\mathcal{E}}} \right)^6 + \left( \frac{\Lambda T^5}{\mathcal{E}} \right) \right]^{-1/6}.
\end{equation}

\textbf{Inertial Limit ($\Lambda \to 0$):} The viscous term vanishes.
\begin{equation}
    V^2 T^2 \approx \mathcal{E} \implies V_{\mathrm{in}} \approx \frac{\sqrt{\mathcal{E}}}{T}.
\end{equation}

\textbf{Viscous Limit ($\Lambda \gg 1$):} The viscous term dominates.
\begin{equation}
    \Lambda V^6 T^5 \approx \mathcal{E} \implies V_{\mathrm{visc}} \approx \left( \frac{\mathcal{E}}{\Lambda T^5} \right)^{1/6}.
\end{equation}

To construct a continuous function across all regimes, we apply a matched asymptotic interpolation. Because the viscous velocity limit is an intrinsic 6th-root ($V_{\mathrm{visc}} \sim \Lambda^{-1/6}$), applying a linear harmonic mean would lead to fractional scaling (e.g., $\Lambda^{1/6}$). To algebraically rationalize the root and preserve the integer unified parameter $\Lambda$ for further reduction when calculating $\beta_{\mathrm{max}}$, we apply a 6th-power harmonic mean ($V^{-6} \approx V_{\mathrm{in}}^{-6} + V_{\mathrm{visc}}^{-6}$):
\begin{equation}
    V \approx \left[ \left( \frac{T}{\sqrt{\mathcal{E}}} \right)^6 + \left( \frac{\Lambda T^5}{\mathcal{E}} \right) \right]^{-1/6}.
\end{equation}
Expanding the inertial term yields:
\begin{equation}
    V \approx \left[ \frac{T^6}{\mathcal{E}^3} + \frac{\Lambda T^5}{\mathcal{E}} \right]^{-1/6}.
\end{equation}
To isolate the initial energy density $\mathcal{E}$, we factor $1/\mathcal{E}^3$ out of the bracket:
\begin{equation}
    V \approx \left( \frac{1}{\mathcal{E}^3} \right)^{-1/6} \left[ T^6 + \frac{\Lambda T^5}{\mathcal{E}} \cdot \mathcal{E}^3 \right]^{-1/6}.
\end{equation}
Simplifying this expression yields the exact closed-form kinematic scaling law for the spreading velocity:
\begin{equation}
    V \approx \sqrt{\mathcal{E}} \left[ T^6 + \Lambda \mathcal{E}^2 T^5 \right]^{-1/6}.
    \label{eq:velocity_exact}
\end{equation}

\begin{figure}[h]
    \centering
    \begin{tikzpicture}
        \begin{axis}[
            width=1.2\figW,
            height=1.0\figH,
            xmode=log,
            ymode=log,
            xmin=1e-5, xmax=1e10,
            ymin=0.003, ymax=2,
            xlabel={$\Lambda \mathcal{E}^2 / T$},
            ylabel={$V/V_{in}$},
            legend pos=south west,
            legend style={at={(0.02,0.40)}, anchor=north west, draw=none, fill=none, font=\footnotesize},
            ticklabel style = {font=\footnotesize}, legend cell align=left
        ]
        
        \addplot [
            dotted, 
            blue, 
            line width=1.0pt
        ] table [x=Gamma, y=closure1, col sep=comma] {SM_Figures/velocity_data.csv};
        \addlegendentry{1st-power Closure}

        \addplot [
            dashed, 
            red, 
            line width=1.2pt
        ] table [x=Gamma, y=closure6, col sep=comma] {SM_Figures/velocity_data.csv};
        \addlegendentry{6th-power Closure}
        
        \addplot [
            solid, 
            black, 
            line width=1.5pt
        ] table [x=Gamma, y=exact, col sep=comma] {SM_Figures/velocity_data.csv};
        \addlegendentry{Exact Implicit Root}

        \end{axis}
    \end{tikzpicture}
    \caption{Comparison of the exact numerical root of the master energy polynomial against analytical approximations for the spreading velocity. The 6th-power closure (red dashed) matches the exact root (black solid) across the entire parameter space. Conversely, attempting a linear 1st-power harmonic mean (blue dotted) introduces significant systematic deviations, mathematically confirming that a 6th-power interpolation is strictly required to rationalize the fractional viscous coupling.}
    \label{fig:velocity_exact_vs_approx}
\end{figure}

\section{S5. Leading-Order Velocity Closure}
\label{sec:velocity_suppression}

In deriving the final unified scaling law (Eq. 5 in the main text), the characteristic velocity within the viscous damping term is evaluated at its leading-order limit, $V \approx \sqrt{\mathcal{E}}$. To justify this mathematically, we substitute the full velocity closure (Eq.~\ref{eq:velocity_exact}) into the viscous term of the timescale interpolation:
\begin{align}
    \left( \Lambda V \mathcal{E}^{3/2} \right)^{1/5} &= \left( \Lambda \left( \sqrt{\mathcal{E}} \big[ T^6 + \Lambda \mathcal{E}^2 T^5 \big]^{-1/6} \right) \mathcal{E}^{3/2} \right)^{1/5} \nonumber \\
    &= (\Lambda \mathcal{E}^2)^{1/5} \big[ T^6 + \Lambda \mathcal{E}^2 T^5 \big]^{-1/30}\\
    &= (\Lambda \mathcal{E}^2)^{1/5}.
\end{align}
Because the dynamic viscous variation is suppressed by the extreme fractional exponent of $-1/30$, the bracketed term has a value between 1 and 1.1 (around 10\% error at its peak) and approaches unity for most of the physical impact parameters. Consequently, the higher-order coupling is mathematically negligible, validating the use of the zeroth-order closure $V \approx \sqrt{\mathcal{E}}$ to achieve a closed analytical form.

\section{S6. Comparison with the Lagubeau et al. (2012) Model}
\label{sec:lagubeau_comparison}

Lagubeau \textit{et al.} \cite{lagubeau2012spreading} proposed an empirical scaling for the spreading time, nondimensionalized by the inertial timescale $D_0/V_0$. The fitted linear equation they proposed is:
\begin{equation}
    t_s / (D_0/V_0) = 0.42 (\Re^{1/10} \We^{1/4}) - 0.25
    \label{eq:Lagubeau}
\end{equation}

While this fit is adequate for the experimental data within their original work, Fig.~\ref{fig:lagubeau comparison 1} demonstrates that it deviates significantly and exhibits massive scatter when tested against more comprehensive modern datasets \cite{aksoy2022spreading,liu2025maximum}. 

\begin{figure}[h]
    \centering
    \input{SM_Figures/Lagubeau_comparison}
    \caption{Comparison of modern, comprehensive literature data \cite{aksoy2022spreading,liu2025maximum} against the empirical power-law fit proposed by Lagubeau \textit{et al.} (2012) \cite{lagubeau2012spreading}. The systematic deviation highlights the physical limitations of utilizing static fractional exponents to capture continuous regime transitions.}
    \label{fig:lagubeau comparison 1}
\end{figure}

This discrepancy arises because the empirical fit in \cite{lagubeau2012spreading} is based on the time the macroscopic optical width $D_w$ stops moving, which is systematically larger than the contact time $t_c$ utilized in this work and by others \cite{aksoy2022spreading,liu2025maximum}. As established in Section S1, this geometric divergence is particularly pronounced at high $\Lambda$, where bulk fluid continues to slump even after the solid-liquid contact line has arrested. 

Fig.~\ref{fig:lagubeau comparison 2} confirms this expected physical behavior. The optical measurements of Lagubeau \textit{et al.} match our theoretical contact-time predictions perfectly at low $\Lambda$, but systematically deviate upwards in the viscous regime where less deformation happens at the contact. While an exact gerometric correlation between the equatorial spreading time and contact spreading time remains an open challenge, when the viscous term of our model is multiplied by a constant geometrical prefactor determined from the data of \cite{lagubeau2012spreading} as 0.65 (order or unity), our proposed equation is capable of capturing the data of \cite{lagubeau2012spreading}. 

\begin{figure}[h]
    \centering
    \pgfplotsset{compat=1.17} 

\definecolor{mycolor0}{RGB}{50,155,105}    
\definecolor{mycolor5}{RGB}{213,81,156}    
\definecolor{mycolor10}{RGB}{54,80,163}    
\definecolor{mycolor125}{RGB}{244,196,48}  
\definecolor{mycolor15}{RGB}{245,127,34}   
\definecolor{mycolorSplash}{RGB}{204,204,255} 
\definecolor{mycolorNo}{RGB}{224,224,224}     

\begin{tikzpicture}
\begin{axis}[
width=1.2\figW,
height=0.9\figH,
at={(0\figW,0\figH)},
    xlabel={\small $\Lambda \mathcal{E}^2$ },
    ylabel={\small $T = t_s/\tau_{\mathrm{ic}}$},
    xmin=0.00001, xmax=10E9,
    ymin=0.003, ymax=2,
    xmode=log,
    ymode=log,
    legend style={at={(0.02,0.47)}, anchor=north west, draw=none, fill=none, font=\footnotesize},
    ticklabel style = {font=\footnotesize}, legend cell align=left
    ]

\addplot[
    only marks, 
    color=blue, 
    mark=o, 
    mark size=1.5pt, 
    line width=0.5pt,
    error bars/.cd, 
        x dir=both, x explicit,
        y dir=both, y explicit,
        error bar style={color=red, thick}
]
table[
    row sep=crcr,
    x index=0, 
    y index=1, 
    x error index=2, 
    y error index=3
]{
10600.10566	0.179949309	298.9084382	0.012809652	\\
32963.50621	0.136712082	614.4100553	0.014335175	\\
28380.85535	0.130504148	651.9573579	0.024471744	\\
117920.2624	0.113680931	6836.358524	0.005607847	\\
250353.9453	0.108986527	7722.411551	0.020324032	\\
20417.69037	0.139302001	862.6639248	0.016329181	\\
44390.29575	0.121895645	911.1461005	0.011853462	\\
7957.130741	0.166864809	834.0142369	0.00874829	\\
20529.11181	0.142861187	695.4056081	0.013572973	\\
5453.568446	0.174865455	121.9165577	0.007406739	\\
2954.709844	0.178513423	101.7886908	0.009508247	\\
5279.588522	0.17129589	175.5346485	0.020533364	\\
2493.654832	0.18572635	166.2351719	0.016114486	\\
1351.505729	0.189470713	72.63553284	0.00696471	\\
2192.108825	0.168066206	46.68119577	0.027897924	\\
661.4619125	0.214563959	14.31281032	0.009267657	\\
274.3435104	0.231168245	10.33140547	0.012647732	\\
275.891613	0.220950138	16.62922109	0.01643229	\\
204.6638486	0.256550506	11.37116939	0.036573988	\\
62.53262276	0.280802564	2.444382985	0.029951519	\\
37.28908254	0.279080055	2.412280905	0.01398768	\\
14.1016738	0.31269322	0.825359393	0.019465425	\\
8.992398929	0.296832194	0.340441962	0.020095398	\\
4.711469298	0.304998997	0.319385982	0.012289286	\\
2.213564982	0.358564277	0.196498014	0.016720148	\\
1.200012952	0.395265948	0.037333625	0.036554127	\\
};
\addlegendentry{Ref.~\cite{lagubeau2012spreading}}

\addplot[line width=2.0pt, black, domain=0.00001:10E11, samples=200] {0.72*(1+x^0.2)^-1};
\addlegendentry{Eq. 4}

\addplot[line width=1.0pt, brown, domain=0.00001:10E11, samples=200] {0.72*(1+0.65*x^0.2)^-1};
\addlegendentry{T= $T_0 [1 + 0.65 (\Lambda \mathcal{E}^2)^{1/5}]^{-1}$}

    \coordinate (A) at (axis cs: 5*1e5, 0.2); 
    
    \coordinate (B) at (axis cs:  5*1e6, 0.2); 
    
    \coordinate (C) at (axis cs:  5*1e6, {0.2 * 10^(-0.2)}); 

    \draw[black, thick] (A) -- (B) node[midway, above] {\scriptsize 1} 
                            -- (C) node[midway, right] {\scriptsize $-\frac{1}{5}$} 
                            -- cycle;

\end{axis}
\end{tikzpicture}
    \caption{Experimental spreading times from Lagubeau \textit{et al.} (2012) \cite{lagubeau2012spreading} based on the equatorial diameter plotted against the theoretical contact-time model $T$ derived in this work. Red error bars are based on 4 repeated measurements.}
    \label{fig:lagubeau comparison 2}
\end{figure}

\section{S7. Comparison with Other Maximum Spreading Models}
\label{sec:maxspread_comparison}

To assess the predictive utility of the proposed closure for practical engineering design, we benchmark the main equation of this work against two widely used equations: the Pad\'e interpolation of Laan et al.~\cite{laan2014maximum}, and the classical energy-balance model of Pasandideh-Fard et al.~\cite{pasandideh1996capillary}.

In practical design and optimization workflows, theoretical models serve as the \textit{a priori} reference to specify operating parameters. A practitioner relies on the theoretical prediction $\beta_{\mathrm{theory}}$ to target a desired droplet footprint; therefore, the relevant design deviation is the fractional departure of the physical outcome from the theoretical expectation:
\begin{equation}
    \text{Deviation Error} = \frac{\beta_{\mathrm{experiment}} - \beta_{\mathrm{theory}}}{\beta_{\mathrm{theory}}}.
    \label{eq:design_error}
\end{equation}

For Laan et al.~\cite{laan2014maximum}, the maximum spreading ratio is given by:
\begin{equation}
    \beta_{\max} = \mathrm{Re}^{1/5} \frac{P^{1/2}}{1.24 + P^{1/2}}, \quad \text{where} \quad P = \mathrm{We}\,\mathrm{Re}^{-2/5},
\end{equation}
and for Pasandideh-Fard et al.~\cite{pasandideh1996capillary}:
\begin{equation}
    \beta_{\max} = \sqrt{\frac{\mathrm{We} + 12}{3(1 - \cos\theta) + 4\left(\dfrac{\mathrm{We}}{\sqrt{\mathrm{Re}}}\right)}}.
\end{equation}
For our model, we evaluate our compact closed-form scaling, supplemented with the geometric correction $\beta_w \approx \frac{1}{4}\beta_c^2 + 1$ (for $\beta_c \le 2$).

\begin{figure}[h!]
    \centering
    \input{SM_Figures/residual}
    \caption{Relative error of the predicted maximum spreading ratio as a function of the unified damping parameter $\Lambda \mathcal{E}^2$ for the present model, Laan et al.~\cite{laan2014maximum}, and Pasandideh-Fard et al.~\cite{pasandideh1996capillary}. The data are from this work, as well as from \cite{liu2025maximum,aksoy2022spreading,mclauchlan2025bouncing,jorgensen2024deformation}. The maximum error for each model is 8.7\% for this work, 258\% for Laan \textit{et al}., and 265\% for Pasandideh-Fard \textit{et al.}.}
    \label{fig:residual}
\end{figure}

Because the residual plot alone may not provide a complete picture, we supplement it with a complementary view of the traditional parity plot in Fig.~\ref{fig:parity}. As can be seen, Laan et al.'s model primarily fails near the deposition limit ($\beta_{\mathrm{exp}} \lesssim 1$), whereas Pasandideh-Fard et al.'s model exhibits a broader range of deviation across the intermediate and viscous regimes.

\begin{figure}[h!]
    \centering
    \begin{subfigure}[c]{0.49\textwidth}
        \centering
        \pgfplotsset{compat=1.17} 

\definecolor{mycolor0}{RGB}{50,155,105}    
\definecolor{mycolor5}{RGB}{213,81,156}    
\definecolor{mycolor10}{RGB}{54,80,163}    
\definecolor{mycolor125}{RGB}{244,196,48}  
\definecolor{mycolor15}{RGB}{245,127,34}   
\definecolor{mycolorSplash}{RGB}{204,204,255} 
\definecolor{mycolorNo}{RGB}{224,224,224}     

\begin{tikzpicture}
\begin{axis}[
width=1.3\figW,
height=1.0\figH,
at={(0\figW,0\figH)},
    xlabel={\small $\beta_{\mathrm{experiment}}$},
    ylabel={\small $\beta_{\mathrm{theory}}$},
    xmin=0.2, xmax=10,
    ymin=0.2, ymax=10,
    xmode=log,
    ymode=log,
    legend style={at={(0.02,0.98)}, anchor=north west, draw=none, fill=none, font=\footnotesize},
    ticklabel style = {font=\footnotesize}, legend cell align=left
    ]

\addplot[line width=1.0pt, black, domain=0.00001:10E11, samples=200] {x};

\addplot[only marks, color=blue, mark=o, mark size=3.0pt, line width=0.1pt]
table[row sep=crcr]{
2.270756757	2.038104911	\\
1.827160494	2.156347465	\\
2.683108108	2.502145994	\\
2.570135135	2.504771181	\\
2.733945946	2.507703194	\\
2.354037267	2.520804434	\\
1.910958904	2.038624564	\\
2.530864198	2.705218687	\\
1.315505913	1.924478836	\\
2.012903226	2.273272085	\\
1.34586071	1.904234804	\\
1.544827586	1.835147804	\\
2.833333333	3.206816953	\\
1.544546649	1.997405458	\\
1.406570302	1.913645535	\\
2.437086093	2.448902177	\\
3.462621622	3.322064191	\\
3.654675676	3.331743364	\\
3.660324324	3.342380013	\\
1.682522996	2.035474531	\\
1.722076216	2.052989074	\\
1.656767411	2.005060461	\\
1.796583443	2.083797312	\\
2.352941176	2.582326858	\\
1.763513514	1.892570707	\\
1.731274639	1.994467353	\\
1.815900131	2.002881191	\\
1.820499343	2.016475078	\\
2.374193548	2.686607482	\\
3.59375	3.647785126	\\
1.850854139	2.039172932	\\
1.310344828	1.480048726	\\
2.807692308	2.770612012	\\
1.998028909	2.144687149	\\
2.007227332	2.221813176	\\
1.248275862	1.367235284	\\
2.065177398	2.191911345	\\
2.093959732	2.04055092	\\
4.05625	4.142856861	\\
2.277660972	2.287830717	\\
1.462585034	1.402505026	\\
1.194444444	1.203439295	\\
3.122580645	3.070264264	\\
2.281879195	2.145253418	\\
1.4	1.251388	\\
4.970810811	4.465636784	\\
4.823945946	4.477347231	\\
4.620594595	4.480013247	\\
1.113636364	1.020019526	\\
1.305555556	1.081712978	\\
4.987341772	4.648733721	\\
1.650684932	1.441156227	\\
3.503225806	3.399534556	\\
2.547297297	2.318847681	\\
1.066176471	0.988615467	\\
1.544827586	1.265763163	\\
1.181818182	0.923949995	\\
1.772413793	1.48695186	\\
5.3	4.952949064	\\
3.698717949	3.601775847	\\
1.416666667	1.085479394	\\
1.627586207	1.302141215	\\
5.620253165	5.148903762	\\
2.378010914	2.010854894	\\
2.34902131	2.01191054	\\
2.368577209	2.013045828	\\
2.365337714	2.022808373	\\
2.362066631	2.028829325	\\
4.044871795	3.748262082	\\
2.80952381	2.520829736	\\
1.092857143	0.717940681	\\
2.440202252	2.05395383	\\
2.439267697	2.059508147	\\
2.442217366	2.066646347	\\
2.457702951	2.074759584	\\
1.265734266	0.916841859	\\
1.951724138	1.576757416	\\
1.496503497	1.110966768	\\
1.784722222	1.374983202	\\
0.927545455	0.716213802	\\
1.019757576	0.716130987	\\
1.019757576	0.716100584	\\
1.321678322	0.935964956	\\
3.19047619	2.743449679	\\
1.126760563	0.699516223	\\
1.608391608	1.170612421	\\
2.160839161	1.69305121	\\
1.090604027	0.57113867	\\
1.103333333	0.570657466	\\
0.98808	0.731256884	\\
0.98092	0.73369298	\\
0.98092	0.73369298	\\
3.356643357	2.880567754	\\
1.958041958	1.472573562	\\
1.001577287	0.551740278	\\
1.154929577	0.71170531	\\
1.00619195	0.516665045	\\
1.408450704	0.984848999	\\
1.008064516	0.47352107	\\
1	0.451706681	\\
1	0.452812879	\\
1	0.445727295	\\
1.004048583	0.443973452	\\
3.61971831	2.98729461	\\
1.004048583	0.425236524	\\
1	0.415720676	\\
1.008733624	0.429408405	\\
1.311778842	0.754304667	\\
1.332335273	0.75440441	\\
1.313427769	0.756010871	\\
1.004716981	0.411553767	\\
1.004694836	0.39745381	\\
1.753521127	1.249601587	\\
2.384615385	1.827859413	\\
2.728323699	2.818511814	\\
2.637724551	2.822449988	\\
2.832891247	2.856272983	\\
2.14893617	1.587306111	\\
1.221428571	0.747325612	\\
1.539007092	1.050257176	\\
1.355138728	0.788905402	\\
1.341418547	0.788972141	\\
1.33070093	0.789412927	\\
1.242151515	0.780126552	\\
1.253	0.780511596	\\
1.280121212	0.78058492	\\
2.560283688	1.921726554	\\
1.1814	0.793400308	\\
1.16708	0.794524833	\\
1.16708	0.795767089	\\
1.0382	0.796742981	\\
0.87352	0.797432457	\\
1.928571429	1.345324462	\\
1.164037855	0.600870236	\\
1.164516129	0.599953611	\\
1.378136755	0.80906359	\\
2.264285714	1.663349704	\\
2.609929078	1.977975048	\\
1.364182439	0.813200516	\\
1.3768032	0.815562774	\\
1.400925106	0.826153626	\\
1.392235743	0.82703192	\\
2.671428571	2.026915433	\\
1.168458781	0.606406306	\\
2.35971223	1.712923725	\\
1.177700348	0.611015851	\\
1.397349832	0.833932007	\\
1.307142857	0.797206123	\\
1.012244898	0.27803629	\\
1.642857143	1.129318	\\
2.021582734	1.407480721	\\
1.413117675	0.840773514	\\
2.731070496	2.094952407	\\
2.431654676	1.753524826	\\
1.445459788	0.848630001	\\
1.442025928	0.849506834	\\
1.418960594	0.849504274	\\
1.456956114	0.84984397	\\
1.426850826	0.850821558	\\
1.008064516	0.277561876	\\
1.366909091	0.847761775	\\
1.361484848	0.848038786	\\
1.350636364	0.848054386	\\
1.473319918	0.861797173	\\
2.650334076	2.196998215	\\
2.587719298	2.201921934	\\
2.420935412	2.200761597	\\
2.116788321	1.452078206	\\
2.661469933	2.202636152	\\
1.478814046	0.865638549	\\
2.641723356	2.203065603	\\
1.463473776	0.865542804	\\
1.496234172	0.867687034	\\
1.507846536	0.870498011	\\
1.003968254	0.277966435	\\
1.016129032	0.278177379	\\
1.499437068	0.876126687	\\
1.762589928	1.182020509	\\
1.511095854	0.88038452	\\
1.518305703	0.884567729	\\
2.176470588	1.491042191	\\
1.518886715	0.887067958	\\
1.405797101	0.856425196	\\
1.288372093	0.782685685	\\
2.294536817	1.781196646	\\
2.33640553	1.792834437	\\
1.300469484	0.786775988	\\
1.276497696	0.789642793	\\
1.290322581	0.790259907	\\
1.288372093	0.789750396	\\
1.791366906	1.220957719	\\
1.00619195	0.296851455	\\
1.27448	0.904086799	\\
2.061946903	1.506070167	\\
1.38188	0.908988456	\\
1.22436	0.910424722	\\
1.00623053	0.299980781	\\
1.328301887	0.830853243	\\
1.846715328	1.246414155	\\
1.334572491	0.835162401	\\
1.335740072	0.839077417	\\
1.389891697	0.839077417	\\
1.389285714	0.841254139	\\
1.384615385	0.838131663	\\
1.319702602	0.836471372	\\
1.364620939	0.840407058	\\
1.335714286	0.842588398	\\
1.012461059	0.300038109	\\
1.529636364	0.916672696	\\
1.507939394	0.917657009	\\
1.012232416	0.303053785	\\
1.545909091	0.918971938	\\
1.474452555	0.893964344	\\
1.012307692	0.306800887	\\
1.013071895	0.301082134	\\
1.00952381	0.313011471	\\
1.009345794	0.314490125	\\
1.019047619	0.314101583	\\
1.00952381	0.314599597	\\
1.722891566	1.206929764	\\
1.4675	0.919360217	\\
1.4875	0.919360217	\\
1.016129032	0.314788869	\\
1.015974441	0.315034351	\\
1.48019802	0.921591749	\\
1.472772277	0.921591749	\\
1.472772277	0.921591749	\\
1.462871287	0.92233476	\\
1.50245098	0.923807226	\\
1.501216545	0.925458459	\\
1.00952381	0.321451036	\\
1.42484	0.961605744	\\
1.009345794	0.323818658	\\
1.496296296	0.924437793	\\
1.016556291	0.316733204	\\
1.01183432	0.315154636	\\
0.723782609	0.386618039	\\
1.5036	0.987114207	\\
1.503816794	0.944890218	\\
0.739347826	0.389464231	\\
0.739347826	0.389676449	\\
1.53224	0.989654469	\\
1.41768	0.990633736	\\
1.019047619	0.328056234	\\
1.461538462	0.958577261	\\
1.275862069	0.707810951	\\
1.611	1.06062914	\\
1.58236	1.061380846	\\
1.611	1.062055369	\\
0.840521739	0.423932588	\\
0.840521739	0.424483775	\\
0.848304348	0.424551815	\\
0.840521739	0.426080227	\\
1.300813008	0.793248985	\\
1.38700565	0.833651816	\\
1.410614525	0.837861919	\\
0.988391304	0.470562128	\\
0.95726087	0.470650013	\\
0.972826087	0.470932938	\\
0.95726087	0.471457291	\\
1.033846154	0.383282097	\\
1.039755352	0.383054276	\\
1.027522936	0.380651319	\\
1.050652174	0.496559794	\\
1.058434783	0.499478932	\\
1.066217391	0.501137267	\\
1.053291536	0.414114569	\\
1.040955631	0.410660748	\\
1.064220183	0.42592438	\\
1.066838046	0.493263492	\\
};
\addlegendentry{Pasandideh-Fard \textit{et al.}~\cite{pasandideh1996capillary}}

\end{axis}
\end{tikzpicture}
        \caption{}
    \end{subfigure}
    \begin{subfigure}[c]{0.49\textwidth}
        \centering
        \pgfplotsset{compat=1.17} 

\definecolor{mycolor0}{RGB}{50,155,105}    
\definecolor{mycolor5}{RGB}{213,81,156}    
\definecolor{mycolor10}{RGB}{54,80,163}    
\definecolor{mycolor125}{RGB}{244,196,48}  
\definecolor{mycolor15}{RGB}{245,127,34}   
\definecolor{mycolorSplash}{RGB}{204,204,255} 
\definecolor{mycolorNo}{RGB}{224,224,224}     

\begin{tikzpicture}
\begin{axis}[
width=1.3\figW,
height=1.0\figH,
at={(0\figW,0\figH)},
    xlabel={\small $\beta_{\mathrm{experiment}}$},
    ylabel={\small $\beta_{\mathrm{theory}}$},
    xmin=0.2, xmax=10,
    ymin=0.2, ymax=10,
    xmode=log,
    ymode=log,
    legend style={at={(0.02,0.98)}, anchor=north west, draw=none, fill=none, font=\footnotesize},
    ticklabel style = {font=\footnotesize}, legend cell align=left
    ]

\addplot[line width=1.0pt, black, domain=0.00001:10E11, samples=200] {x};

\addplot[only marks, color=teal, mark=triangle, mark size=3.0pt, line width=0.1pt]
table[row sep=crcr]{
2.270756757	0.756794994	\\
1.827160494	1.20993211	\\
2.683108108	1.869614995	\\
2.570135135	1.873658721	\\
2.733945946	1.878165168	\\
2.354037267	1.907920107	\\
1.910958904	1.126391584	\\
2.530864198	2.173183432	\\
1.315505913	1.183034661	\\
2.012903226	1.759865007	\\
1.34586071	1.20756776	\\
1.544827586	1.036058236	\\
2.833333333	2.811372531	\\
1.544546649	1.523231671	\\
1.406570302	1.40724868	\\
2.437086093	2.082119034	\\
3.462621622	2.934507615	\\
3.654675676	2.946021158	\\
3.660324324	2.958664335	\\
1.682522996	1.640632762	\\
1.722076216	1.689351602	\\
1.656767411	1.641734041	\\
1.796583443	1.769439629	\\
2.352941176	2.301625252	\\
1.763513514	1.53954966	\\
1.731274639	1.694159859	\\
1.815900131	1.718065258	\\
1.820499343	1.755383547	\\
2.374193548	2.464998122	\\
3.59375	3.339073366	\\
1.850854139	1.814674725	\\
1.310344828	0.913292855	\\
2.807692308	2.593181849	\\
1.998028909	1.989292947	\\
2.007227332	2.076908276	\\
1.248275862	0.833824179	\\
2.065177398	2.089286021	\\
2.093959732	1.963214309	\\
4.05625	3.930484185	\\
2.277660972	2.278632165	\\
1.462585034	1.278618668	\\
1.194444444	0.791355611	\\
3.122580645	3.036020792	\\
2.281879195	2.190139101	\\
1.4	1.167657999	\\
4.970810811	4.289807747	\\
4.823945946	4.303922149	\\
4.620594595	4.307136411	\\
1.113636364	0.762902511	\\
1.305555556	1.078592531	\\
4.987341772	4.545419372	\\
1.650684932	1.567908413	\\
3.503225806	3.509342509	\\
2.547297297	2.521251026	\\
1.066176471	0.506822687	\\
1.544827586	1.423501939	\\
1.181818182	0.966004331	\\
1.772413793	1.720572724	\\
5.3	4.920276407	\\
3.698717949	3.795809881	\\
1.416666667	1.297273453	\\
1.627586207	1.568868069	\\
5.620253165	5.162901892	\\
2.378010914	2.341402082	\\
2.34902131	2.343994694	\\
2.368577209	2.345525096	\\
2.365337714	2.364202197	\\
2.362066631	2.375790707	\\
4.044871795	4.001286657	\\
2.80952381	2.870268029	\\
1.092857143	0.801434098	\\
2.440202252	2.423878361	\\
2.439267697	2.433582148	\\
2.442217366	2.44743436	\\
2.457702951	2.462841253	\\
1.265734266	1.159046439	\\
1.951724138	1.945533659	\\
1.496503497	1.413365147	\\
1.784722222	1.768005811	\\
0.927545455	0.929695255	\\
1.019757576	0.933934891	\\
1.019757576	0.936472041	\\
1.321678322	1.260261563	\\
3.19047619	3.228245204	\\
1.126760563	0.96261666	\\
1.608391608	1.588133647	\\
2.160839161	2.183891766	\\
1.090604027	0.787196312	\\
1.103333333	0.792750593	\\
0.98808	1.0397393	\\
0.98092	1.05286854	\\
0.98092	1.05286854	\\
3.356643357	3.438372959	\\
1.958041958	1.978154156	\\
1.001577287	0.279559453	\\
1.154929577	1.043835389	\\
1.00619195	0.296188067	\\
1.408450704	1.412378792	\\
1.008064516	0.289112496	\\
1	0.298988586	\\
1	0.296595474	\\
1	0.299450778	\\
1.004048583	0.300195117	\\
3.61971831	3.597270546	\\
1.004048583	0.309104032	\\
1	0.314182612	\\
1.008733624	0.301857206	\\
1.311778842	1.133588875	\\
1.332335273	1.133776362	\\
1.313427769	1.138810549	\\
1.004716981	0.307426447	\\
1.004694836	0.314717255	\\
1.753521127	1.764862074	\\
2.384615385	2.426660294	\\
2.728323699	3.461185124	\\
2.637724551	3.468765309	\\
2.832891247	3.503473591	\\
2.14893617	2.191504296	\\
1.221428571	1.164238757	\\
1.539007092	1.56324047	\\
1.355138728	1.229925564	\\
1.341418547	1.230483908	\\
1.33070093	1.231770332	\\
1.242151515	1.220121226	\\
1.253	1.22109032	\\
1.280121212	1.221274637	\\
2.560283688	2.583060375	\\
1.1814	1.243017681	\\
1.16708	1.24575495	\\
1.16708	1.248761952	\\
1.0382	1.251112091	\\
0.87352	1.252766196	\\
1.928571429	1.94770048	\\
1.164037855	0.98896353	\\
1.164516129	0.987798196	\\
1.378136755	1.278578659	\\
2.264285714	2.32159305	\\
2.609929078	2.673130099	\\
1.364182439	1.287851084	\\
1.3768032	1.292986897	\\
1.400925106	1.316565528	\\
1.392235743	1.318412493	\\
2.671428571	2.749619889	\\
1.168458781	1.014791529	\\
2.35971223	2.402946904	\\
1.177700348	1.023354332	\\
1.397349832	1.333208472	\\
1.307142857	1.284129644	\\
1.012244898	0.48715435	\\
1.642857143	1.717584288	\\
2.021582734	2.056446514	\\
1.413117675	1.347939259	\\
2.731070496	2.848669546	\\
2.431654676	2.467889241	\\
1.445459788	1.364281194	\\
1.442025928	1.365662339	\\
1.418960594	1.365777405	\\
1.456956114	1.366608714	\\
1.426850826	1.368235472	\\
1.008064516	0.499261574	\\
1.366909091	1.369773462	\\
1.361484848	1.370328115	\\
1.350636364	1.370359341	\\
1.473319918	1.390221219	\\
2.650334076	2.977080589	\\
2.587719298	2.982884642	\\
2.420935412	2.982034081	\\
2.116788321	2.131135281	\\
2.661469933	2.984500247	\\
1.478814046	1.397485404	\\
2.641723356	2.985848705	\\
1.463473776	1.397664066	\\
1.496234172	1.40163814	\\
1.507846536	1.406692693	\\
1.003968254	0.516011541	\\
1.016129032	0.518178762	\\
1.499437068	1.418077372	\\
1.762589928	1.811557412	\\
1.511095854	1.426360921	\\
1.518305703	1.434383074	\\
2.176470588	2.194536823	\\
1.518886715	1.438916324	\\
1.405797101	1.40332509	\\
1.288372093	1.304848091	\\
2.294536817	2.543805539	\\
2.33640553	2.558654777	\\
1.300469484	1.311508481	\\
1.276497696	1.31578041	\\
1.290322581	1.316975362	\\
1.288372093	1.316627353	\\
1.791366906	1.877904967	\\
1.00619195	0.571308004	\\
1.27448	1.471652165	\\
2.061946903	2.226383182	\\
1.38188	1.480627889	\\
1.22436	1.483246249	\\
1.00623053	0.583073591	\\
1.328301887	1.380634216	\\
1.846715328	1.920138557	\\
1.334572491	1.38767251	\\
1.335740072	1.393041682	\\
1.389891697	1.393041682	\\
1.389285714	1.39637451	\\
1.384615385	1.392223722	\\
1.319702602	1.390095563	\\
1.364620939	1.395498708	\\
1.335714286	1.398835752	\\
1.012461059	0.586920193	\\
1.529636364	1.499569272	\\
1.507939394	1.501328063	\\
1.012232416	0.593750219	\\
1.545909091	1.503674233	\\
1.474452555	1.472353391	\\
1.012307692	0.604791045	\\
1.013071895	0.595505859	\\
1.00952381	0.618130376	\\
1.009345794	0.621233697	\\
1.019047619	0.620972506	\\
1.00952381	0.622013498	\\
1.722891566	1.886874735	\\
1.4675	1.514668204	\\
1.4875	1.514668204	\\
1.016129032	0.622946626	\\
1.015974441	0.623457809	\\
1.48019802	1.517980873	\\
1.472772277	1.517980873	\\
1.472772277	1.517980873	\\
1.462871287	1.519276026	\\
1.50245098	1.52126723	\\
1.501216545	1.523715011	\\
1.00952381	0.635661692	\\
1.42484	1.573570579	\\
1.009345794	0.640505658	\\
1.496296296	1.525938052	\\
1.016556291	0.630848235	\\
1.01183432	0.62859304	\\
0.723782609	0.749602016	\\
1.5036	1.616786216	\\
1.503816794	1.560921099	\\
0.739347826	0.756329796	\\
0.739347826	0.75682633	\\
1.53224	1.621034824	\\
1.41768	1.622670144	\\
1.019047619	0.657054043	\\
1.461538462	1.59165121	\\
1.275862069	1.253420791	\\
1.611	1.736371137	\\
1.58236	1.73756194	\\
1.611	1.738629987	\\
0.840521739	0.83051747	\\
0.840521739	0.831626551	\\
0.848304348	0.831763333	\\
0.840521739	0.834828782	\\
1.300813008	1.382158127	\\
1.38700565	1.444980055	\\
1.410614525	1.451304105	\\
0.988391304	0.919427363	\\
0.95726087	0.919587618	\\
0.972826087	0.920103373	\\
0.95726087	0.921058648	\\
1.033846154	0.77958236	\\
1.039755352	0.779414465	\\
1.027522936	0.775479725	\\
1.050652174	0.965978673	\\
1.058434783	0.971109668	\\
1.066217391	0.974016817	\\
1.053291536	0.838568045	\\
1.040955631	0.83333609	\\
1.064220183	0.858918366	\\
1.066838046	0.971793665	\\
};
\addlegendentry{Laan \textit{et al.} ~\cite{laan2014maximum}}

\end{axis}
\end{tikzpicture}
        \caption{}
    \end{subfigure}
    \begin{subfigure}[c]{0.49\textwidth}
        \centering
        \pgfplotsset{compat=1.17} 

\definecolor{mycolor0}{RGB}{50,155,105}    
\definecolor{mycolor5}{RGB}{213,81,156}    
\definecolor{mycolor10}{RGB}{54,80,163}    
\definecolor{mycolor125}{RGB}{244,196,48}  
\definecolor{mycolor15}{RGB}{245,127,34}   
\definecolor{mycolorSplash}{RGB}{204,204,255} 
\definecolor{mycolorNo}{RGB}{224,224,224}     

\begin{tikzpicture}
\begin{axis}[
width=1.3\figW,
height=1.0\figH,
at={(0\figW,0\figH)},
    xlabel={\small $\beta_{\mathrm{experiment}}$},
    ylabel={\small $\beta_{\mathrm{theory}}$},
    xmin=0.2, xmax=10,
    ymin=0.2, ymax=10,
    xmode=log,
    ymode=log,
    legend style={at={(0.02,0.98)}, anchor=north west, draw=none, fill=none, font=\footnotesize},
    ticklabel style = {font=\footnotesize}, legend cell align=left
    ]

\addplot[only marks, color=red, mark=square, mark size=3.0pt, line width=0.1pt]
table[row sep=crcr]{
2.270756757	2.754812491	\\
1.827160494	2.151761392	\\
2.683108108	2.516672499	\\
2.570135135	2.519329278	\\
2.733945946	2.522296168	\\
2.354037267	2.53200824	\\
1.910958904	1.927528251	\\
2.530864198	2.719897523	\\
1.315505913	1.81121822	\\
2.012903226	2.259951703	\\
1.34586071	1.802396011	\\
1.544827586	1.695652734	\\
2.833333333	3.235610751	\\
1.544546649	1.978335821	\\
1.406570302	1.878675654	\\
2.437086093	2.489463971	\\
3.462621622	3.350686233	\\
3.654675676	3.360739733	\\
3.660324324	3.371796079	\\
1.682522996	2.052118821	\\
1.722076216	2.084505427	\\
1.656767411	2.034890632	\\
1.796583443	2.139774505	\\
2.352941176	2.662160091	\\
1.763513514	1.924153558	\\
1.731274639	2.05482474	\\
1.815900131	2.071435911	\\
1.820499343	2.097807044	\\
2.374193548	2.797241943	\\
3.59375	3.703450409	\\
1.850854139	2.140758782	\\
1.310344828	1.463047239	\\
2.807692308	2.906375313	\\
1.998028909	2.287450025	\\
2.007227332	2.372000315	\\
1.248275862	1.391400455	\\
2.065177398	2.367060625	\\
2.093959732	2.228450251	\\
4.05625	4.249759052	\\
2.277660972	2.524162291	\\
1.462585034	1.59286829	\\
1.194444444	1.326526182	\\
3.122580645	3.298837238	\\
2.281879195	2.414972467	\\
1.4	1.493829512	\\
4.970810811	4.595036692	\\
4.823945946	4.608362822	\\
4.620594595	4.611398098	\\
1.113636364	1.270722077	\\
1.305555556	1.407491834	\\
4.987341772	4.829528157	\\
1.650684932	1.759954738	\\
3.503225806	3.734791631	\\
2.547297297	2.704431653	\\
1.066176471	1.174477302	\\
1.544827586	1.62447828	\\
1.181818182	1.322872222	\\
1.772413793	1.870920324	\\
5.3	5.185822183	\\
3.698717949	4.003232137	\\
1.416666667	1.508812763	\\
1.627586207	1.719041444	\\
5.620253165	5.417112586	\\
2.378010914	2.483152823	\\
2.34902131	2.485412442	\\
2.368577209	2.486887413	\\
2.365337714	2.50377225	\\
2.362066631	2.514248551	\\
4.044871795	4.19711868	\\
2.80952381	3.023005944	\\
1.092857143	1.220007385	\\
2.440202252	2.557835455	\\
2.439267697	2.566773904	\\
2.442217366	2.57933122	\\
2.457702951	2.593354671	\\
1.265734266	1.401082021	\\
1.951724138	2.060296641	\\
1.496503497	1.576588888	\\
1.784722222	1.870867223	\\
0.927545455	1.263534878	\\
1.019757576	1.26505426	\\
1.019757576	1.265970436	\\
1.321678322	1.453228081	\\
3.19047619	3.357733273	\\
1.126760563	1.272187382	\\
1.608391608	1.695596934	\\
2.160839161	2.27615842	\\
1.090604027	1.187629645	\\
1.103333333	1.189254791	\\
0.98808	1.30872455	\\
0.98092	1.314573573	\\
0.98092	1.314573573	\\
3.356643357	3.556514016	\\
1.958041958	2.056167382	\\
1.001577287	1.053797476	\\
1.154929577	1.306323186	\\
1.00619195	1.054688255	\\
1.408450704	1.544890508	\\
1.008064516	1.049672395	\\
1	1.049934592	\\
1	1.049540554	\\
1	1.049538286	\\
1.004048583	1.049543807	\\
3.61971831	3.707639697	\\
1.004048583	1.049752476	\\
1	1.049936193	\\
1.008733624	1.048677604	\\
1.311778842	1.353346825	\\
1.332335273	1.353451084	\\
1.313427769	1.356031962	\\
1.004716981	1.048280791	\\
1.004694836	1.048480611	\\
1.753521127	1.835082079	\\
2.384615385	2.502511838	\\
2.728323699	3.561285968	\\
2.637724551	3.568260834	\\
2.832891247	3.60362179	\\
2.14893617	2.25492857	\\
1.221428571	1.365755682	\\
1.539007092	1.65024001	\\
1.355138728	1.405519241	\\
1.341418547	1.405805879	\\
1.33070093	1.40652695	\\
1.242151515	1.398918661	\\
1.253	1.399468288	\\
1.280121212	1.399572891	\\
2.560283688	2.650522274	\\
1.1814	1.412891514	\\
1.16708	1.414481468	\\
1.16708	1.416233495	\\
1.0382	1.417606726	\\
0.87352	1.418575312	\\
1.928571429	1.99814812	\\
1.164037855	1.262813715	\\
1.164516129	1.262192104	\\
1.378136755	1.434035161	\\
2.264285714	2.378034917	\\
2.609929078	2.736324998	\\
1.364182439	1.439654579	\\
1.3768032	1.442798905	\\
1.400925106	1.457336291	\\
1.392235743	1.458496034	\\
2.671428571	2.809460994	\\
1.168458781	1.274005324	\\
2.35971223	2.455557067	\\
1.177700348	1.278295659	\\
1.397349832	1.467814559	\\
1.307142857	1.434409032	\\
1.012244898	1.067820719	\\
1.642857143	1.771713236	\\
2.021582734	2.101200947	\\
1.413117675	1.477185719	\\
2.731070496	2.905207097	\\
2.431654676	2.517686502	\\
1.445459788	1.487754393	\\
1.442025928	1.488698221	\\
1.418960594	1.488760904	\\
1.456956114	1.48928969	\\
1.426850826	1.49039154	\\
1.008064516	1.070083844	\\
1.366909091	1.490592855	\\
1.361484848	1.490957233	\\
1.350636364	1.490977752	\\
1.473319918	1.504929122	\\
2.650334076	3.032359023	\\
2.587719298	3.038171754	\\
2.420935412	3.037221143	\\
2.116788321	2.172495958	\\
2.661469933	3.039641931	\\
1.478814046	1.509826763	\\
2.641723356	3.040831497	\\
1.463473776	1.509907103	\\
1.496234172	1.512606139	\\
1.507846536	1.516067334	\\
1.003968254	1.073421617	\\
1.016129032	1.073880248	\\
1.499437068	1.523756521	\\
1.762589928	1.852112098	\\
1.511095854	1.529423082	\\
1.518305703	1.534953534	\\
2.176470588	2.233287804	\\
1.518886715	1.538120962	\\
1.405797101	1.51117926	\\
1.288372093	1.442210087	\\
2.294536817	2.585622397	\\
2.33640553	2.600386632	\\
1.300469484	1.446544132	\\
1.276497696	1.449384165	\\
1.290322581	1.450142702	\\
1.288372093	1.449854936	\\
1.791366906	1.91173062	\\
1.00619195	1.087420334	\\
1.27448	1.56105109	\\
2.061946903	2.262696933	\\
1.38188	1.567474469	\\
1.22436	1.569356603	\\
1.00623053	1.090446643	\\
1.328301887	1.49308718	\\
1.846715328	1.950896066	\\
1.334572491	1.497915201	\\
1.335740072	1.501744126	\\
1.389891697	1.501744126	\\
1.389285714	1.504075189	\\
1.384615385	1.501087412	\\
1.319702602	1.499546496	\\
1.364620939	1.503404964	\\
1.335714286	1.505743173	\\
1.012461059	1.09132199	\\
1.529636364	1.580574388	\\
1.507939394	1.581857535	\\
1.012232416	1.093273859	\\
1.545909091	1.583571828	\\
1.474452555	1.559296628	\\
1.012307692	1.096323862	\\
1.013071895	1.09341855	\\
1.00952381	1.100302986	\\
1.009345794	1.101244706	\\
1.019047619	1.101130118	\\
1.00952381	1.101447113	\\
1.722891566	1.915129456	\\
1.4675	1.590386122	\\
1.4875	1.590386122	\\
1.016129032	1.101697432	\\
1.015974441	1.101853604	\\
1.48019802	1.592907722	\\
1.472772277	1.592907722	\\
1.472772277	1.592907722	\\
1.462871287	1.593866932	\\
1.50245098	1.595414781	\\
1.501216545	1.597285668	\\
1.00952381	1.105691867	\\
1.42484	1.636563755	\\
1.009345794	1.107210265	\\
1.496296296	1.598439875	\\
1.016556291	1.103877975	\\
1.01183432	1.103116837	\\
0.723782609	1.145606796	\\
1.5036	1.670267568	\\
1.503816794	1.624826585	\\
0.739347826	1.148008777	\\
0.739347826	1.14818718	\\
1.53224	1.673634601	\\
1.41768	1.674933143	\\
1.019047619	1.111958544	\\
1.461538462	1.647521631	\\
1.275862069	1.401054364	\\
1.611	1.768667788	\\
1.58236	1.76968527	\\
1.611	1.770598492	\\
0.840521739	1.176316721	\\
0.840521739	1.176764351	\\
0.848304348	1.176819606	\\
0.840521739	1.178060707	\\
1.300813008	1.486292727	\\
1.38700565	1.530471668	\\
1.410614525	1.535050891	\\
0.988391304	1.214368412	\\
0.95726087	1.214440882	\\
0.972826087	1.214674207	\\
0.95726087	1.215106745	\\
1.033846154	1.154199829	\\
1.039755352	1.15411631	\\
1.027522936	1.152566816	\\
1.050652174	1.235993468	\\
1.058434783	1.238447039	\\
1.066217391	1.239843327	\\
1.053291536	1.177581751	\\
1.040955631	1.175341473	\\
1.064220183	1.186189734	\\
1.066838046	1.237749906	\\
};
\addlegendentry{This work}

\addplot[line width=1.0pt, black, domain=0.00001:10E11, samples=200] {x};

\end{axis}
\end{tikzpicture}
        \caption{}
    \end{subfigure}
    \caption{Parity plot comparing experimental equatorial maximum spreading ratios $\beta_{\mathrm{exp}}$ against model predictions $\beta_{\mathrm{model}}$ for the present work, Laan et al.~\cite{laan2014maximum}, and Pasandideh-Fard et al.~\cite{pasandideh1996capillary}. The solid line indicates $y = x$. The data are from this work, as well as from \cite{liu2025maximum,aksoy2022spreading,mclauchlan2025bouncing,jorgensen2024deformation}.}
    \label{fig:parity}
\end{figure}

Furthermore, to quantitatively compare the three models, we compute the root-mean-square (RMS) error across the regimes in Table~\ref{tab:rms_comparison}. 

\begin{table}[]
\centering
\caption{Root-mean-square (RMS) error comparison for $\beta_w$ across different regimes of the unified damping parameter $\Lambda\mathcal{E}^2$.}
\label{tab:rms_comparison}
\begin{tabular}{lccc}
\hline\hline
Regime & Present & Laan et al. & Pasandideh-Fard et al. \\
\hline
Overall & 0.21 & 0.29 & 0.49 \\
$\Lambda\mathcal{E}^2 < 10$ & 0.29 & 0.70 & 0.31 \\
$\Lambda\mathcal{E}^2 \ge 10$ & 0.21 & 0.27 & 0.49 \\
\hline\hline
\end{tabular}
\end{table}

\clearpage

\bibliographystyle{apsrev4-2} 
\bibliography{references}